\documentclass[12pt]{article}
\usepackage{cite}
\usepackage{graphics}
\usepackage{epsfig}

\newcommand{\GeV}{\mbox{ GeV}}
\newcommand{\TeV}{\mbox{ TeV}}

\newcommand{\stopp}{\ensuremath{\tilde t}}
\newcommand{\sbottom}{\ensuremath{\tilde b}} 
\newcommand{\stau}{\ensuremath{\tilde \tau}}
\newcommand{\sneut}{\ensuremath{\tilde \nu}}
\newcommand{\sg}{\ensuremath{\tilde{g}}}

\newcommand{\cplus}{\ensuremath{\chi^+}}

\newcommand{\neut}{\ensuremath{\chi^0}}

\newcommand{\gsim}{\mbox{ \raisebox{-4pt}{${\stackrel{\textstyle >}{\sim}}$} }}
\newcommand{\lsim}{\mbox{ \raisebox{-4pt}{${\stackrel{\textstyle <}{\sim}}$} }}
\newcommand{\tb}{\ensuremath{\tan\beta}}

\newcommand{\mb}{\ensuremath{m_b}}
\newcommand{\mt}{\ensuremath{m_t}}
\newcommand{\mHp}{\ensuremath{M_{H^\pm}}}
\newcommand{\mA}{\ensuremath{M_{A^0}}}
\newcommand{\mg}{\ensuremath{m_{\tilde{g}}}}

\newcommand{\msbo}{\ensuremath{m_{\tilde{b}_1}}}

\newcommand{\msto}{\ensuremath{m_{\tilde{t}_1}}}
\newcommand{\mstt}{\ensuremath{m_{\tilde{t}_2}}}
\newcommand{\msta}{\ensuremath{m_{\tilde{t}_a}}}

\newcommand{\tch}{\ensuremath{t\rightarrow c\,h}}

\newcommand{\figtbw}{
\begin{figure}[t]
\begin{center}
\resizebox{7cm}{!}{\includegraphics{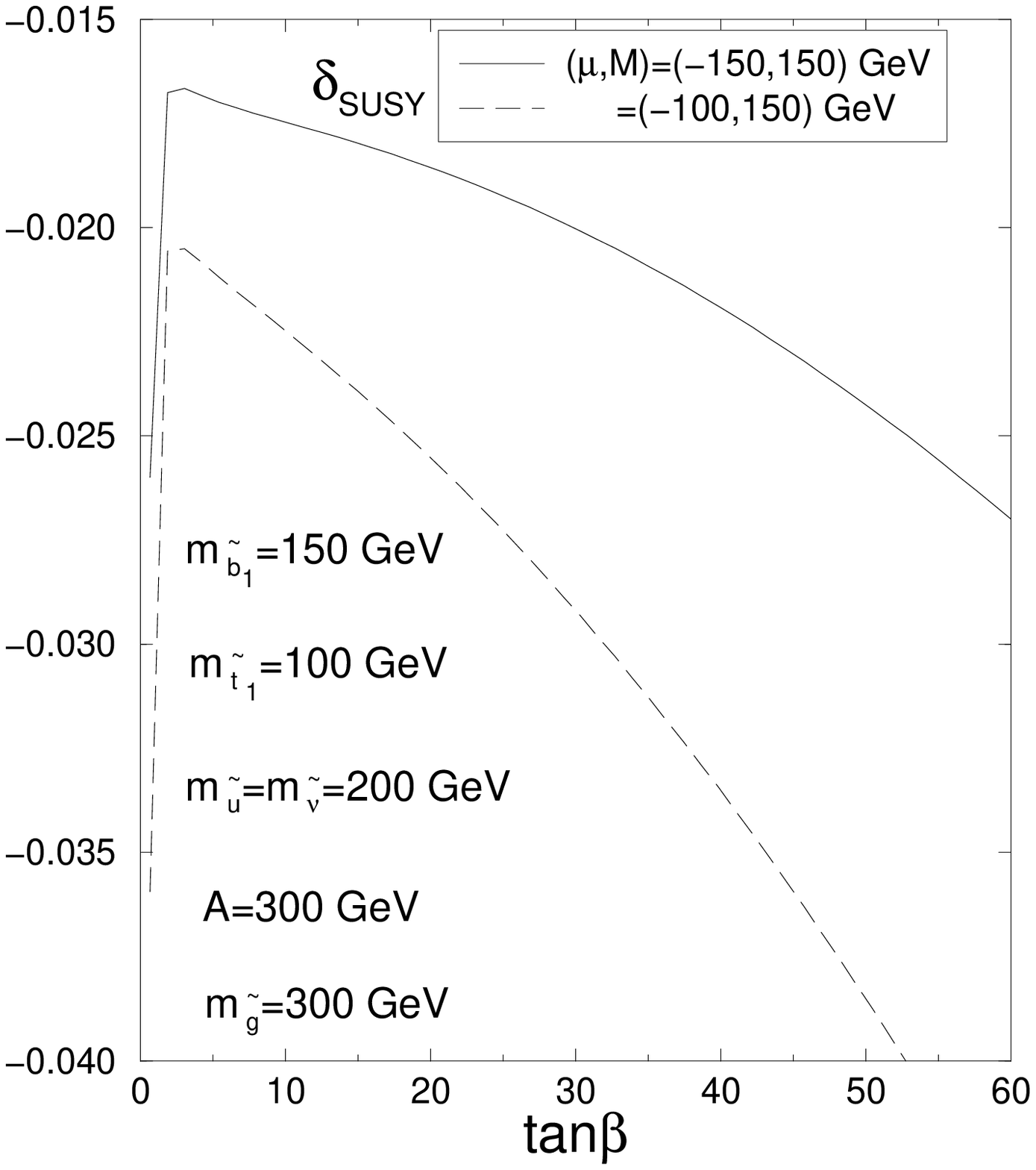}}
\end{center}
\vspace*{-.8cm}
\caption{\sf  The total (electroweak and strong) SUSY correction to
$\Gamma (t\rightarrow W^+\,b)$ for given sets of parameters and $\mt=175\GeV$.
 \label{fig:tbw}} 
\end{figure}
}

\newcommand{\figtbh}{
\begin{figure}[t]
\begin{center}
\begin{tabular}{cc}
\resizebox{7.5cm}{!}{\includegraphics{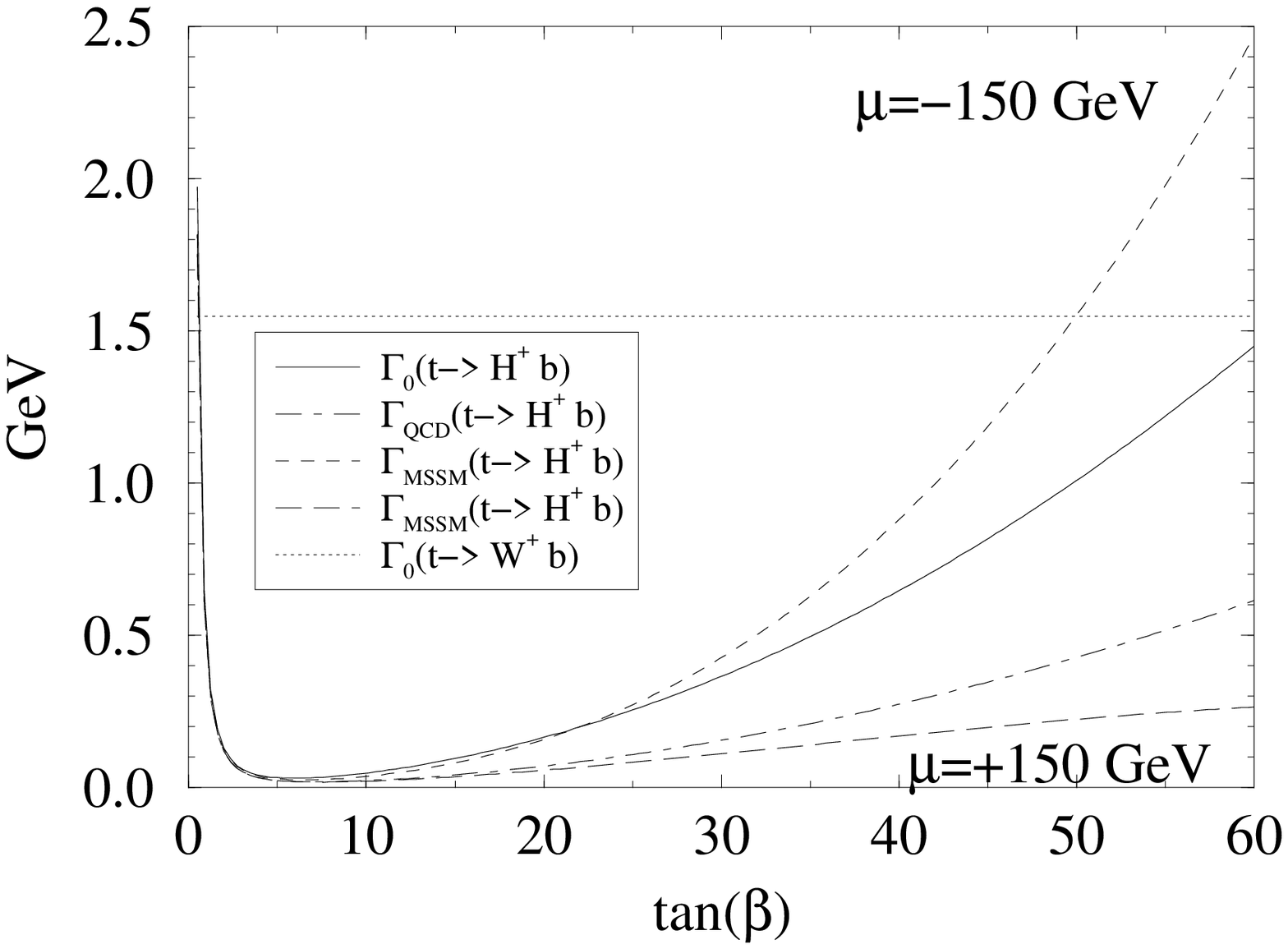}} & 
\resizebox{7.5cm}{!}{\includegraphics{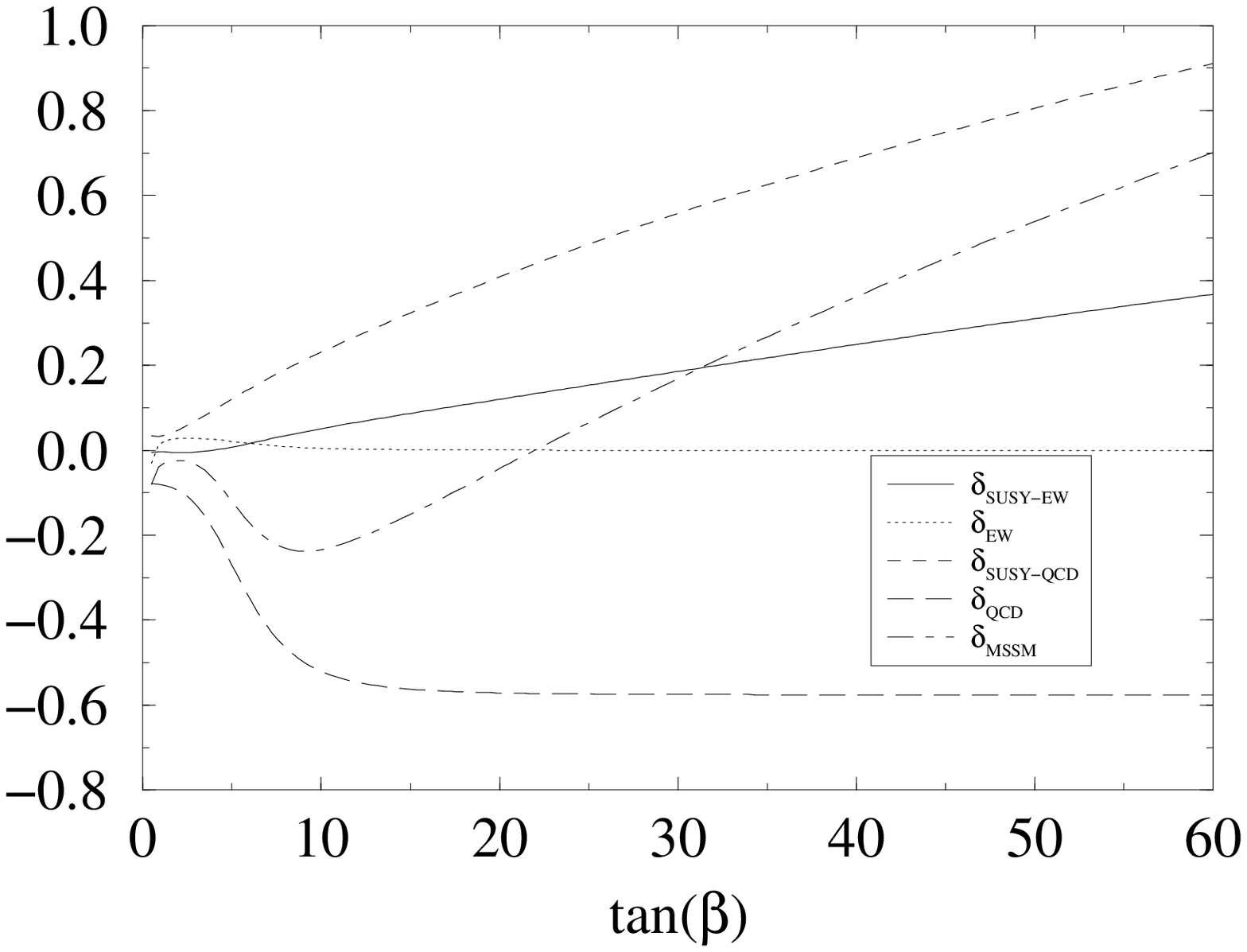}} \\ 
(a) &(b)
\end{tabular}
\end{center}
\vspace*{-.8cm}
\caption{\sf \textbf{(a)} The top-quark partial decay  width $\Gamma(t\to H^+b)$
  compared with the standard one as a function of \tb, and for
  $\mHp=120\GeV$. Shown are the tree-level 
  width, the QCD corrected width, and the full MSSM corrected width for two
  sets of the SUSY parameters \textit{A}: \{$\mu,\msto,\msbo,\mg,A_t$\}=
                                           $\{-150,100,150,300,+300\}$ GeV
\textit{B}:$\{+150,200,600,1000,-300\}$ GeV. 
  \textbf{(b)} The relative radiative corrections to $\Gamma(t\to H^+b)$ for
  each of the sectors of the MSSM (\textit{A} parameter set).\label{fig:tbh}} 
\end{figure}
}

\newcommand{\figtch}{
\begin{figure}[t]
\begin{center}
\begin{tabular}{cc}
\resizebox{7.5cm}{!}{\includegraphics{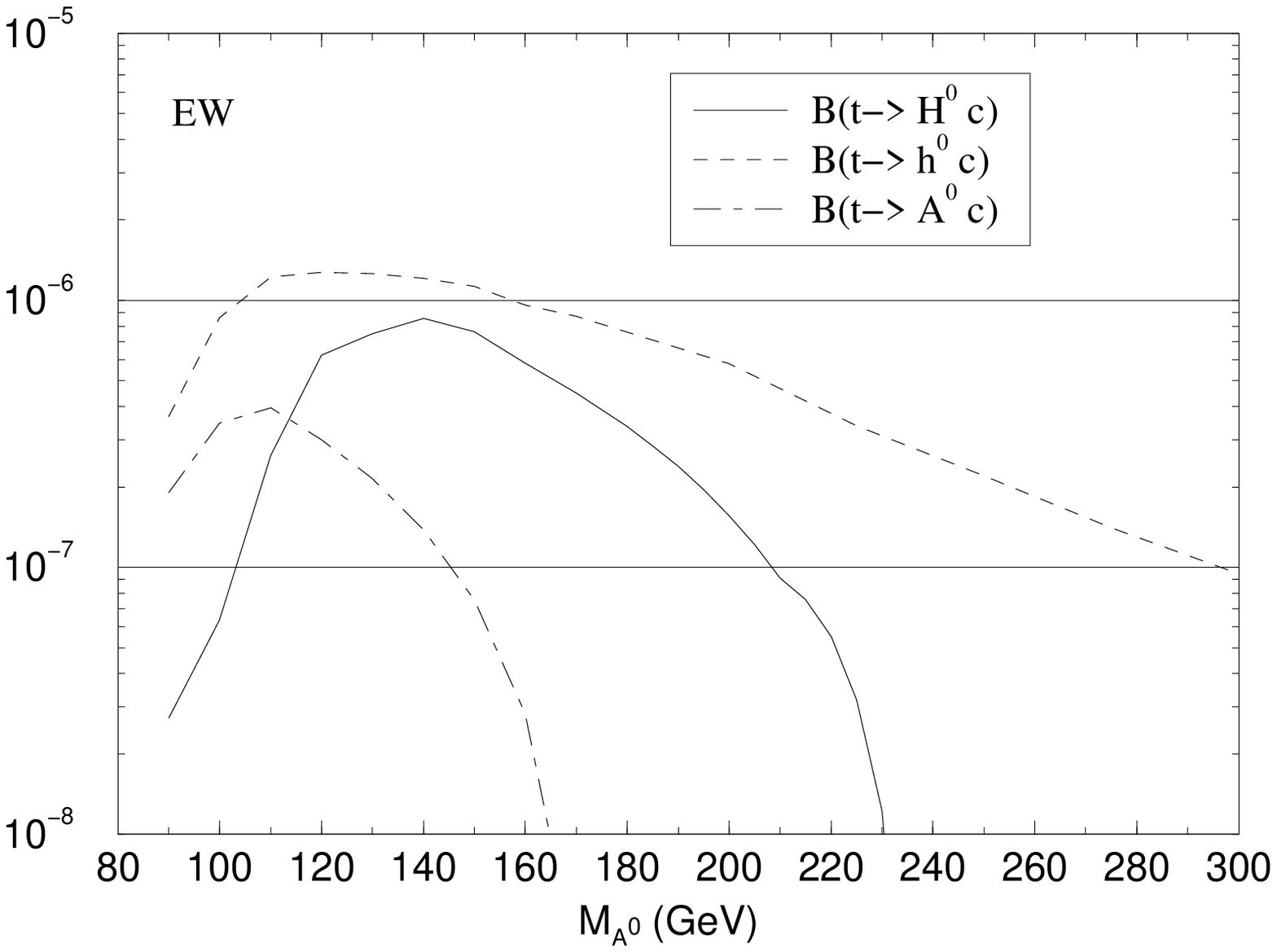}} & 
\resizebox{7.5cm}{!}{\includegraphics{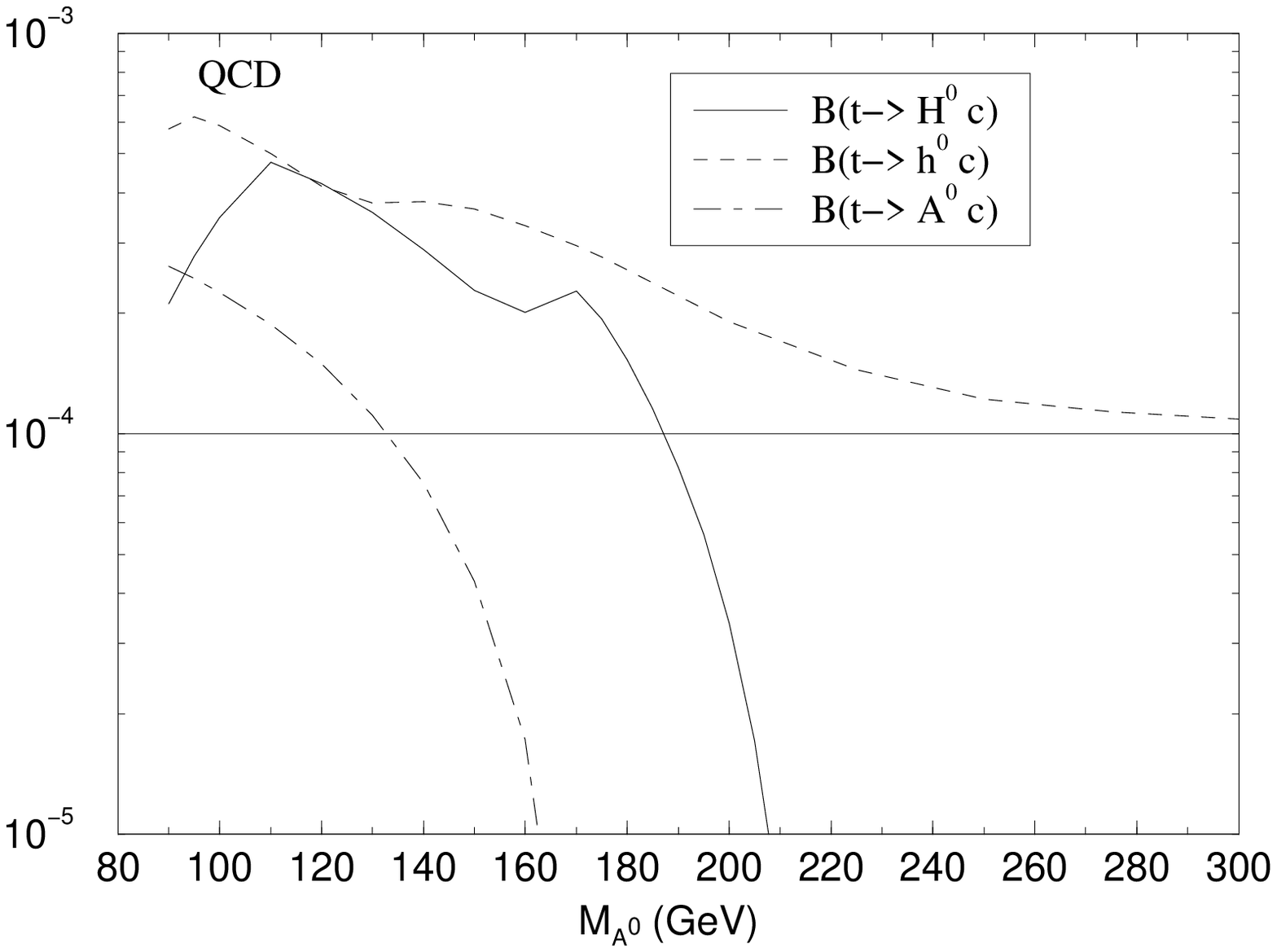}} \\ 
(a) & (b) \\ 
\multicolumn{2}{c}{\resizebox{7.5cm}{!}{\includegraphics{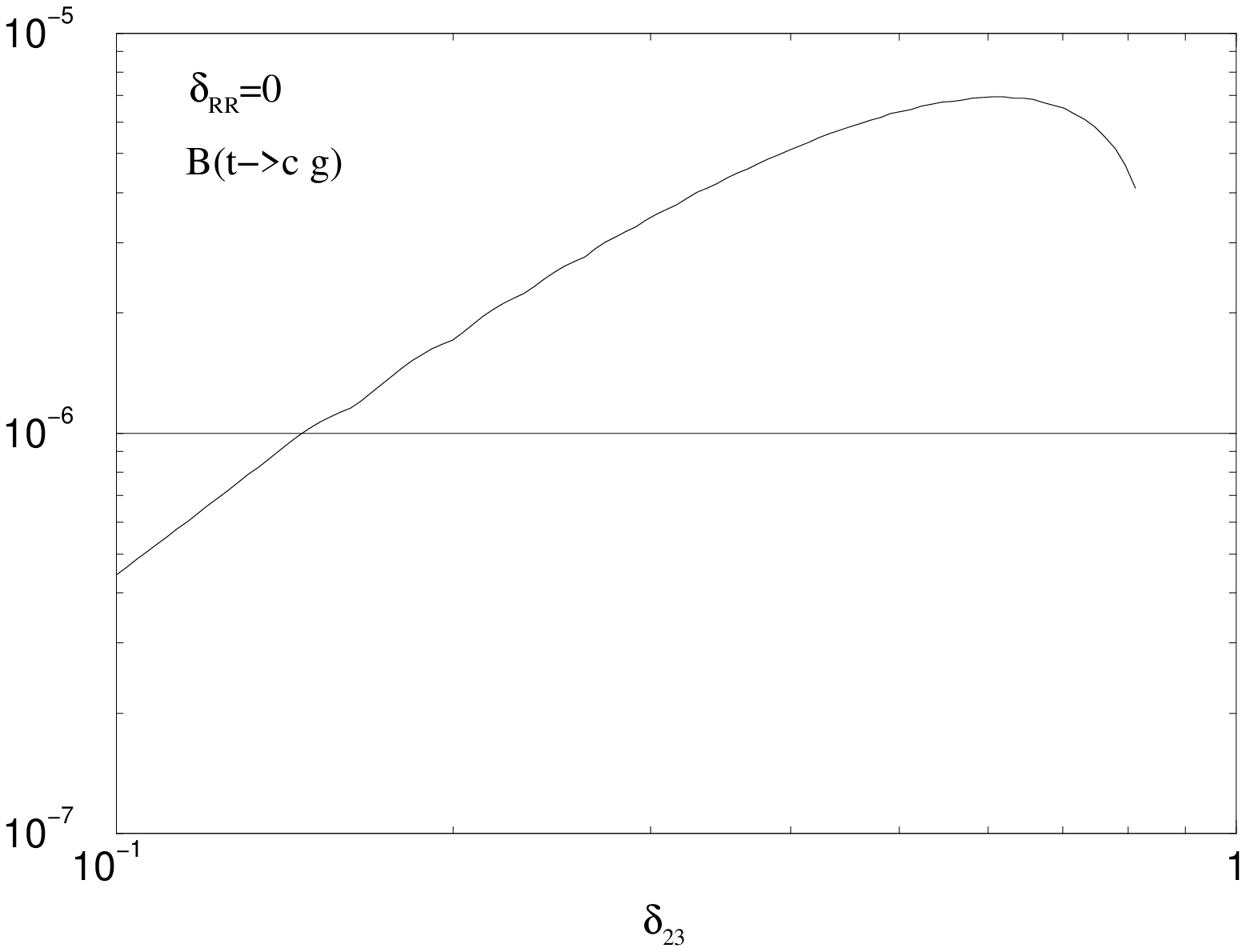}}} \\ 
\multicolumn{2}{c}{(c)}
\end{tabular}
\end{center}
\vspace*{-.8cm}
\caption{\sf \textbf{(a)} Maximum value of $B(t\rightarrow c\,h)$, obtained by
taking into account only the SUSY-EW contributions, as a function of $\mA$ ; 
\textbf{(b)} as in (a) but taking into account only the SUSY-QCD
contributions; and \textbf{(c)} maximum value of $B(t\rightarrow c\,g)$ as a
function of the intergenerational mixing parameter $\delta _{23}$ in the LH
sector. In all cases the scanning for the rest of parameters of the MSSM has
been performed within the phenomenologically allowed region.\label{fig:tch}} 
\end{figure}
}

\newcommand{\figtstneut}{
\begin{figure}[t]
\begin{center}
\begin{tabular}{cc}
\includegraphics[width=7.8cm,clip]{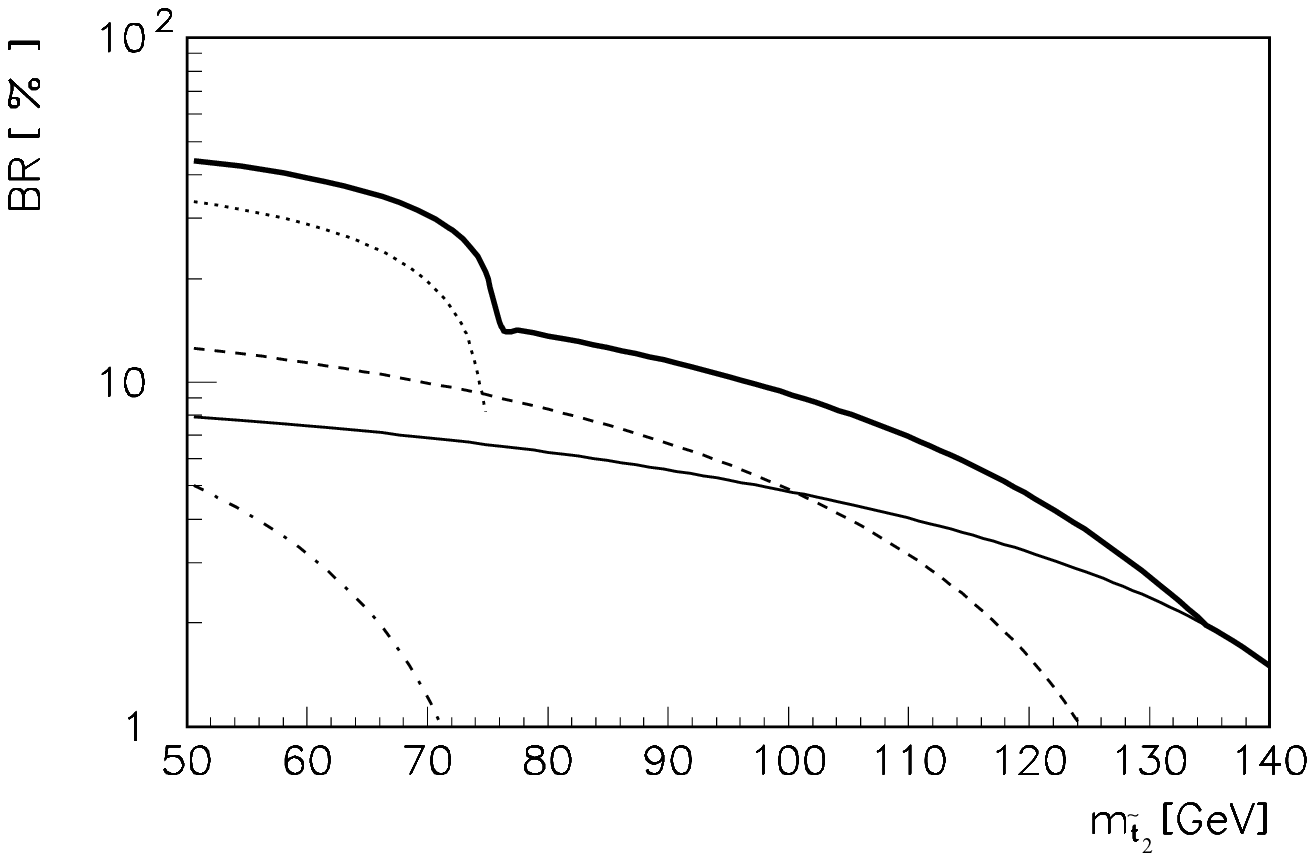} &
\includegraphics[width=8cm,clip]{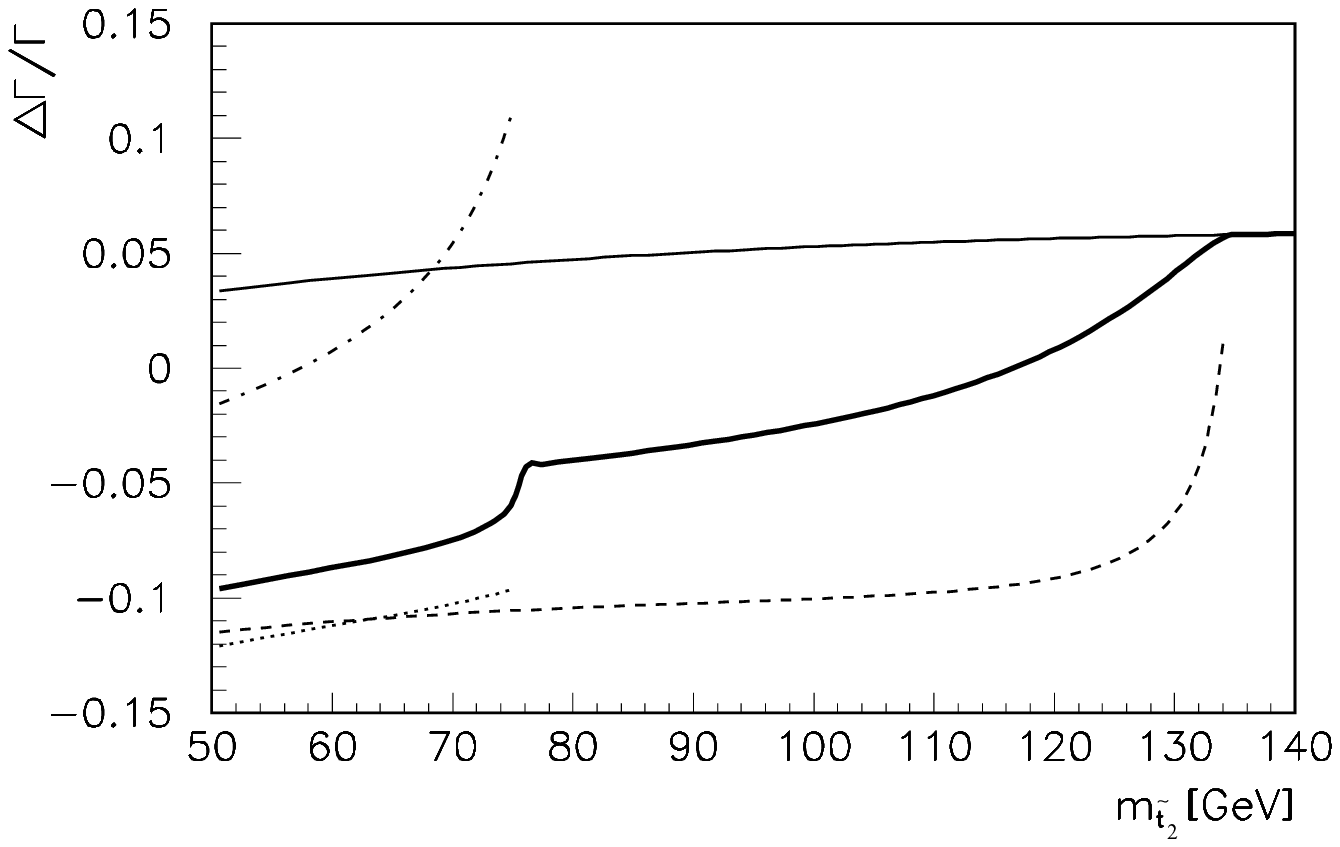} \\
(a) &(b)
\vspace*{-.8cm}
\end{tabular}
\end{center}
\caption{\sf \textbf{(a)} The tree-level prediction for $\mbox{BR}(t\to\stopp\neut_\alpha)$
  as a function of the lightest stop mass. \textbf{(b)} The one-loop SUSY-QCD
  corrections to $\Gamma(t\to\stopp\neut_\alpha)$. $M_2=-\mu=50\GeV$, $\tb=1.6$
  $\mt=180\GeV$. The thin, solid, dashed and dash-dotted lines correspond
  respectively to
 $\alpha=1-4$, the thick solid line corresponds to
  the sum over 
  all neutralinos. $\mg$($\simeq 180\GeV$) is fixed from GUT relations. From Ref.~\cite{Djouadi:1996pi}.\label{fig:tstneut}}  
\end{figure}
}

\newcommand{\bea}{\begin{equation}}
\newcommand{\eea}{\end{equation}}
\newcommand{\beq}{\begin{eqnarray}}
\newcommand{\eeq}{\end{eqnarray}}
\newcommand{\ba}{\begin{array}}
\newcommand{\ea}{\end{array}}
\newcommand{\dd}{{\rm d}}
\newcommand{\dfrac}{\displaystyle\frac}
\newcommand{\nn}{\nonumber}
\newcommand{\gmu}{\gamma_\mu}

\newcommand{\gfi}{\gamma_5}

\newcommand{\smunu}{\sigma_{\mu\nu}}

\newcommand{\eps}{\epsilon}

\newcommand{\gev}{{\rm GeV}}

\def\plb#1#2#3{{\it Phys. Lett. }{\bf B#1~}(19#2)~#3}

\def\prl#1#2#3{{\it Phys. Rev. Lett. }{\bf #1~}(19#2)~#3}

\def\prd#1#2#3{{\it Phys. Rev. }{\bf D#1~}(19#2)~#3}
\def\npb#1#2#3{{\it Nucl. Phys. }{\bf B#1~}(19#2)~#3}

\def\sjnp#1#2#3{{\it Sov. J. Nucl. Phys. }{\bf #1~}(19#2)~#3}

\def\pplus{{\mathbf{\hat p}_{\mathbf{+}}}}

\def\kplus{{\mathbf{\hat k}_{\mathbf{+}}}}

\def\qplus{{\mathbf{q}_{\mathbf{+}}}}
\def\qminus{{\mathbf{q}_{\mathbf{-}}}}

\def\sone{{\mathbf{s}_{\mathbf{1}}}}
\def\stwo{{\mathbf{s}_{\mathbf{2}}}}
\def\soner{{\mathbf{s}^{\mathbf{*}}_{\mathbf{1}}}}
\def\stwor{{\mathbf{s}^{\mathbf{*}}_{\mathbf{2}}}}

\def\ppmn{{\mathbf{p}_{\mathbf{\pm}}}}
\def\pmpn{{\mathbf{p}_{\mathbf{\mp}}}}
\def\kplusn{{\mathbf{k}_{\mathbf{+}}}}

\def\kpmn{{\mathbf{k}_{\mathbf{\pm}}}}
\def\kmpn{{\mathbf{k}_{\mathbf{\mp}}}}
\def\qplusn{{\mathbf{\hat q}_{\mathbf{+}}}}
\def\qminusn{{\mathbf{\hat q}_{\mathbf{-}}}}
\def\qpmn{{\mathbf{q}_{\mathbf{\pm}}}}
\def\qmpn{{\mathbf{q}_{\mathbf{\mp}}}}
\begin{document}
\thispagestyle{empty}

\begin{flushright}
{\parbox{3.5cm}{
DESY 00-036\\
KA-TP-4-2000\\
UAB-FT-483\\
UG-FT-113/00\\ 
LC-TH-2000-030\\
hep-ph/0003109\\
}}
\end{flushright}

\vspace{0.5cm}

\begin{center}

{\large\sc {\bf Top-Quark Production and Decay in the
    MSSM}}\footnote{Contribution to the proceedings of
         the 2nd Joint ECFA/DESY Workshop on 
          {\it Physics and Detectors for a Linear Electron-Positron
          Collider}. Update of the talks presented by J.I. Illana at Frascati
    (Italy) 
    November 8-10th 1998, J. Guasch at Oxford (UK) March 20-23th 1999,
    J. Sol{\`a} at 
    Sitges (Spain) April 28th-May 5th 1999 and
    W. Hollik at  Obernai (France) October 16-19th  1999.}

\vspace{0.8cm}

{\sc 
  J.~Guasch$^{a}$%
, W.~Hollik$^{a}$%
, J.I.~Illana$^{b,c}$%
, C.~Schappacher$^{a}$%
, J.~Sol\`a$^{d}$%
}

\vspace*{0.8cm}

{\sl

$^a$ Institut f\"ur Theoretische Physik, Universit\"at Karlsruhe,  D-76128 Karlsruhe, Germany

$^b$ DESY, Platanenallee 6, D-15738 Zeuthen, Germany

$^c$ Dpto. F{\'\i}sica Te\'orica y del Cosmos, Universidad de Granada,
     E-18071 Granada, Spain

$^d$  Grup de F{\'\i}sica Te{\`o}rica and Institut de F{\'\i}sica d'Altes
Energies, Universitat Aut{\`o}noma de Barcelona, E-08193 Bellaterra (Barcelona), Catalonia, Spain

}

\end{center}

\vspace*{0.5cm}

\begin{abstract} 
We review the features of  top-quark decays and loop-induced
effects in the production cross section
and CP-violating observables 
of $e^+e^- \rightarrow t\bar{t}$ which are specific to the 
$R$-parity conserving Minimal Supersymmetric Standard Model (MSSM). 
\end{abstract}

\newpage

\section{Introduction}
The Standard Model (SM) of strong (QCD) and electroweak (EW) interactions has
been the most successful framework to describe the phenomenology of high energy
physics. However, one of the fundamental building blocks (the Higgs boson) still
lacks experimental confirmation. On the other hand, the SM still suffers from some
theoretical deficiencies, most notably the hierarchy problem. Several extensions of
the SM have been proposed to solve these problems; in this note we will
concentrate on its supersymmetric (SUSY) extension, more specifically to the
$R$-parity conserving Minimal Supersymmetric Standard Model (MSSM)~\cite{MSSM}.

The top quark, due to its large mass, could play a central role in the search of
physics beyond the SM. On one hand it could decay to non-standard particles, on
the other hand, due to its large Yukawa coupling, the effects of the
Spontaneous Symmetry Breaking Sector are expected to be larger than for any other
particle of the model. In the MSSM, these effects are reinforced by the presence
of the SUSY partners of the top quark and Higgs bosons. Moreover, the new parameters
appearing in the MSSM can have complex phases, and new sources of CP-violation
phenomena can appear.

One should also bear in mind that, before the commissioning of TESLA, the LHC will
be producing data which might turn out to include some physics beyond the SM,
and we must be prepared for whatever the high energy physics scenario would be
for the running of TESLA.

Here we present a review of the effects of SUSY in the top-quark
phenomenology. First we review in section~\ref{sec:decay} the top-quark decay
within the framework of the MSSM; then, in section~\ref{sec:prod}, we present
the MSSM effects on the top quark pair production in $e^+e^-$ collisions, both
in the presence and absence of CP-violating couplings.

\section{Top-quark decays}
\label{sec:decay}

The existence of SUSY could affect the total top-quark decay
width in two ways. First of all through
unexpected radiative corrections to the standard top quark
decay process $t\rightarrow W^{+}b$. Second, some of the SUSY particles could be
lighter than the top quark itself, thus 
providing new channels in which the top quark could decay.

Concerning the standard top-quark decay, we point out that
its branching ratio $\mbox{BR}(t\rightarrow W^{+}\,b)$ is not so severely
constrained by the observed top quark production cross section as one could
naively think at first sight~\cite{D0EPS99}. In the MSSM the total observed
cross section can be written, schematically, as
\begin{equation}%
\begin{array}
[c]{rcl}%
\sigma_{\mathrm{obs}} & = & \int dq\,d\bar{q}\,\,\sigma(q\,\bar{q}\rightarrow
t\,\bar{t})\,\times|\mbox{BR}(t\rightarrow W^{+}\,b)|^{2}\nonumber\\
& + & \int dq\,d\bar{q}\,\,\sigma(q\,\bar{q}\rightarrow\tilde{g}\,\bar
{\tilde{g}})\,\times|\mbox{BR}(\tilde{g}\rightarrow t\,\bar{\tilde{t}}_{1}%
)|^{2}\times|\mbox{BR}(t\rightarrow W^{+}\,b)|^{2}\nonumber\\
& + & \int dq\,d\bar{q}\,\,\sigma(q\,\bar{q}\rightarrow\tilde{b}_{a}%
\,\bar{\tilde{b}}_{a})\,\times|\mbox{BR}(\tilde{b}_{a}\rightarrow t\,\chi_{1}%
^{-})|^{2}\times|\mbox{BR}(t\rightarrow W^{+}\,b)|^{2}+\ldots\,
\end{array}
\label{eq:productionMSSM}%
\end{equation}
Here $\int dq$ represents the integration over the quarks Parton
Distribution Functions, and sumation over quark flavours. In the SM only the
first line of this equation is present. 
It follows that, in the MSSM, our present ignorance of the SUSY parameters
prevents us from performing a detailed calculation of the
$t\bar{t}$ production cross section as well as from putting a strict
limit on $\mbox{BR}(t\rightarrow$``new''$)$  --- the branching ratio
of the top quark into new physics. It should also be clear that the
observed cross section in Eqn.~(\ref{eq:productionMSSM}) refers not only to the
standard $bW\,bW$ events, but to all kind of final states that can mimic
them. Thus, effectively, we should substitute $\mbox{BR}(t\rightarrow X\,b)$ in that
formula for $\mbox{BR}(t\rightarrow W^{+}\,b)$, and then sum the cross section over
$X$, where $X$ is any state that leads to an observed pattern of leptons and
jets similar to those resulting from $W$-decay. In particular, $X=H^{\pm}$
would contribute (see below) to the $\tau$-lepton signature, if \tb\  is large
enough. Similarly, there can be direct top quark decays into SUSY particles
that could mimic the SM decay of the top quark~\cite{Guasch:1997dk}.
The only restriction is an approximate lower bound $\mbox{BR}(t\rightarrow W^{+}\,b)\gsim40-50\%$ in order to guarantee the
purported standard top quark events at the Tevatron~\cite{CDFD0}. 
Thus, from these considerations it is not excluded that the non-SM branching
ratio of the top quark, $\mbox{BR}(t\rightarrow$``new''$)$, could be
comparable to the SM one --- or at least not necessarily much smaller.

The $e^{+}e^{-}$ Linear Collider (LC) provides an excellent tool to test the
various top quark partial decays widths. The total top quark decay width can
be measured by means of a threshold scan, in a model independent
way~\cite{teubner}. 
Also the clean environment allows for a high
prospect  for detecting exclusive rare decay channels.

\subsection{The top-quark standard decay width}

In the SM the top-quark decays into a $W^{+}$ gauge boson and a bottom quark.
The tree-level prediction for this partial decay width  is
(for $m_{t}=175\,GeV$)
\begin{eqnarray}
\Gamma_{\mathrm{SM}}^{0}(t\rightarrow W^{+}b) & = & 
\left(  {\displaystyle\frac{G_{F}}%
{8\pi\sqrt{2}}}\right)  
{\displaystyle\frac{|V_{tb}|^{2}}{m_{t}}}\lambda^{1/2}%
(1,{m_{b}^{2}}/{m_{t}^{2}},{M_{W}^{2}}/{m_{t}^{2}})\nonumber\\
&  & \times\lbrack M_{W}^{2}(m_{t}^{2}+m_{b}^{2})+(m_{t}^{2}-m_{b}^{2}%
)^{2}-2M_{W}^{4}]\simeq1.55\GeV\,\,
\label{eq:treetbw}%
\end{eqnarray}
where $\lambda^{1/2}(1,x^{2},y^{2})$ is the usual K\"{a}llen function.
As the width turns out to be much larger than the typical
scale of non-perturbative QCD effects $\Lambda_{QCD}$, it can
be conceived as an effective infrared cut-off. This means that the top quark,
due to its large mass, has time to weakly decay before 
strong hadronization processes come into play. And
for this reason perturbative computations in top quark
physics are reliable.

In this spirit the SM quantum corrections to the standard top quark decay
 width have been performed. The 
short-distance QCD effects have been computed up to
two-loop
level~\cite{tdecaySMQCD,tdecaySMEilam}
and they amount to  a correction of $-10\%$ with
respect to the tree-level width. The electroweak (EW) SM
radiative correction is also available~\cite{tdecaySMEilam,tdecaySMEW},
but this contribution is below $+2\%$ in a scheme where
the tree-level width is parametrized in terms of the Fermi constant
$G_{F}$ --- as in Eqn.(\ref{eq:treetbw}). In this scheme the
electroweak corrections are minimized both in the SM and in the MSSM because
the set of universal contributions --- viz.\  those encoded in the parameter
$\Delta r$ --- cancel out.

The MSSM may furnish extra (perturbative) quantum effects on
the standard decay width of the top quark through the one-loop corrections
mediated by non-SM particles. They have been computed
in~\cite{Garcia:1994rq,Dabelstein:1995jt}\footnote{The corresponding
corrections in the general 2HDM can be found in
Refs.~\cite{Grzadkowski:1992nj,Denner:1993vz}. The
particularization of these Higgs effects in the MSSM is studied in~\cite{Grzadkowski:1992nj}.} and can be of two types,
electroweak and strong. The SUSY-EW quantum corrections~\cite{Garcia:1994rq}
are negative (as the standard QCD ones) and vary from $-1\%$
to $-10\%$, depending on the choice of the various SUSY parameters, and,
especially of \tb. The corrections due to 
additional Higgs particle exchange (i.e.\  the MSSM Higgs effects after
subtracting the corresponding SM limit of the MSSM Higgs sector) are at most of
$0.1\%$ due to the severe 
constraints that SUSY imposes on the MSSM Higgs sector. 

\figtbw

The SUSY-QCD corrections, mediated by gluinos and squarks, have also been
found to be negative in most of the parameter space~\cite{Dabelstein:1995jt},
 though they are in general smaller than the SUSY-EW
ones --- namely, around a few \% level --- and they are independent of
\tb.

In Fig.~\ref{fig:tbw} we show the total SUSY (electroweak and QCD) corrections
to this decay width for typical values of the SUSY spectrum.
 The upshot is that the total SUSY corrections to
$\Gamma(t\rightarrow W^{+}\,b)$ go in the same direction as the standard QCD
ones. For large values of  
$\tan\beta$ they can typically yield an effect about half the size of
the QCD corrections, thus providing an additional (potentially measurable)
decrease of the tree-level value of the standard width~(\ref{eq:treetbw}).

\subsection{Top-quark decay into charged Higgs}

\label{sec:tbh} If the charged Higgs is light enough the top quark will also
decay through the process $t\rightarrow H^{+}b$. This decay has been subject
of interest since very early in the
literature~\cite{tbHold}. 
If \tb\  is large or small enough the tree-level prediction for the partial
decay width $\Gamma^{0}(t\rightarrow H^{+}b)$ is comparable to the standard
one~(\ref{eq:treetbw}). In fact $\Gamma^{0}(t\rightarrow H^{+}b)$ presents a
minimum at the point $\tb=\sqrt{\mt/\mb}\simeq6$ and grows
for larger or smaller values of \tb\  (see Fig.~\ref{fig:tbh}). Remarkably
enough, this process turns out to be extremely 
sensitive to radiative corrections of all kinds. On one hand, the standard QCD
corrections are quite large. They are negative~and for $\tb
\gsim10$ they saturate around the value
$-58\%$~\cite{Czarnecki}. 
On the other hand,
the full set of the MSSM radiative corrections, at the one-loop
level, can also be very important. They have been computed
in~\cite{Yang:1994ra,Guasch:1995rn,Coarasa:1996qa}, and more recently the
general 
2HDM corrections became also available~\cite{Coarasa:1998xy}.

In the case of the EW corrections one needs to define
renormalization prescriptions also for the non-SM parameters that
appear in these decays, in particular for the
highly relevant parameter \tb. The renormalization counterterm for
\tb\  can be fixed in many different ways. The actual corrections will
depend on the particular definition, but not so the value of the physical
observable, of course. In our case we found it practical to define \tb\  through
the condition that the partial decay width $\Gamma(H^{+}\rightarrow\tau
^{+}\nu_{\tau})$ does not receive radiative corrections. This is a good choice
for the scenario under study, since this is the dominant decay of a light
charged Higgs boson (i.e. $\mHp<\mt$) provided $\tb$ is of order $1$ or above:
$\tb>\sqrt{m_{c}/m_{s}}\gsim2$. 
Under this renormalization prescription, and for moderate or large $\tb
\gsim10$, the bulk of the SUSY quantum corrections are known to stem  from the finite threshold corrections
to the bottom mass counterterm. The relevant effects are
triggered by $R$-odd particles entering the bottom self-energy, and
can be cast as follows~\cite{Guasch:1995rn,Coarasa:1996qa} 
\begin{eqnarray}
\left(  {\displaystyle\frac{\delta m_{b}}{m_{b}}}\right)  _{\mathrm{S-QCD}}&=&
{\displaystyle\frac
{2\alpha_{s}(m_{t})}{3\pi}}\,m_{\tilde{g}}\,M_{LR}^{b}\,I(m_{\tilde{b}_{1}%
},m_{\tilde{b}_{2}},m_{\tilde{g}})\nonumber\\ &\rightarrow&-{\displaystyle
\frac{2\alpha_{s}(m_{t})}{3\pi
}}\,m_{\tilde{g}}\,\mu\tb\,I(m_{\tilde{b}_{1}},m_{\tilde{b}_{2}},m_{\tilde{g}%
})\,,\nonumber\\
\left(  {
\displaystyle\frac{\delta m_{b}}{m_{b}}}\right) _{\mathrm{S-EW}}&=&-{
\displaystyle\frac
{h_{t}\,h_{b}}{16\pi^{2}}}\,\,{
\displaystyle\frac{\mu}{m_{b}}}\,m_{t}\,M_{LR}%
^{t}I(m_{\tilde{t}_{1}},m_{\tilde{t}_{2}},\mu)\nonumber\\ &\rightarrow&-{
\displaystyle\frac{h_{t}^{2}%
}{16\pi^{2}}}\,\mu\,A_{t}\tb\,I(m_{\tilde{t}_{1}},m_{\tilde{t}_{2}},\mu)\,,\nonumber\\
I(m_{1},m_{2},m_{3})&\equiv&16\,\pi^{2}i\,C_{0}(0,0,m_{1},m_{2},m_{3}%
)\nonumber\\ &=&{\displaystyle\frac{m_{1}^{2}\,m_{2}^{2}\ln{\displaystyle\frac{m_{1}^{2}}{m_{2}^{2}}}+m_{2}^{2}%
\,m_{3}^{2}\ln{\displaystyle\frac{m_{2}^{2}}{m_{3}^{2}}}+m_{1}^{2}\,m_{3}^{2}\ln
{\displaystyle\frac{m_{3}^{2}}{m_{1}^{2}}}}{(m_{1}^{2}-m_{2}^{2})\,(m_{2}^{2}-m_{3}%
^{2})\,(m_{1}^{2}-m_{3}^{2})}}\,\,,\label{eq:dmb}%
\end{eqnarray}
where $C_{0}$ is the three-point 't Hooft-Passarino-Veltman
function~\cite{PV}, 
and the rightmost expressions
hold for sufficiently large \tb. Several important consequences can be derived
already from the approximate expressions~(\ref{eq:dmb}). The first one
corresponds to the sign of the corrections. The sign of the
SUSY-QCD corrections is opposite to that of the higgsino mass parameter $\mu$,
whereas the sign of the SUSY-EW corrections is given by the
product of $\mu$ and the soft-SUSY-breaking trilinear coupling $A_{t}$.
Second, both kind of corrections grow linearly with \tb. A third, and
important, observation is that, if we scale all the dimensionful parameters of
Eqn.~(\ref{eq:dmb}) by a factor $\lambda$, the $\lambda$-dependence drops out
in the final expression. This means that raising the scale of SUSY breaking
does not reduce the effects of the radiative corrections. Notice, however, that this consideration amounts to scale up also
the trilinear coupling $A_{t}$ as well as the higgsino parameter
$\mu$. Therefore, it may lead to unwanted fine-tuning effects in at
least two important sectors of the MSSM: in the Higgs and squark sectors.
Notwithstanding, if one does not stretch out the ranges of the parameters up
to unreasonable limits (i.e. much beyond $1$ TeV or so), an important
consequence can be derived without disrupting the natural structure of the
model, to wit: that the $R$-odd particles whose masses are above the
EW scale can effectively display, in the presence of Yukawa couplings, a
non-decoupling behaviour. And this behaviour is triggered by the existence of
explicit soft SUSY-breaking terms in combination with the spontaneous breaking
of the gauge symmetry. Obviously, this is a very important feature as it could
produce visible radiative corrections for this decay.\footnote{We should like to
  say that this feature is not limited to just the high energy 
process under consideration, but it applies equally well to some low-energy
processes, e.g. in B-meson decays~\cite{Coarasa:1997uw}.} Recently these finite threshold
effects have been further refined in the literature and they have been
re-summed to all orders~\cite{Carena:1999py}. 

In the case of the gluino mass dependence there is another trait that
we wish to remark. Even without scaling up the rest of the SUSY parameters,
the SUSY-QCD corrections to $\Gamma^{0}(t\rightarrow H^{+}b)$ exhibit
a local, and lengthy sustained,  maximum around $\mg\gsim300\GeV$.
Only for gluino masses well above the TeV scale (for fixed values of the
squark masses) do these corrections eventually decouple~\cite{Guasch:1995rn}.
As for the corrections due to Higgs bosons loops, of which there are
quite a few, we find that they are entirely negligible compared to the yield
from $R$-odd particles.\footnote{This effect is due to the
restrictions that SUSY imposes to the form of the Higgs bosons potential.
In the unrestricted 2HDM the Higgs bosons loop corrections
can also be important~\cite{Coarasa:1998xy}.} 

\figtbh

Figure~\ref{fig:tbh}a presents a summary of the main results. Here we have
plotted the partial decay width $\Gamma(t\rightarrow H^{+}b)$ as a function of
\tb, for the two different scenarios that have been identified. In the first
one ($\mu<0$) the SUSY-QCD corrections are opposite in sign to the QCD ones,
thus canceling partially (or even totally) the SM strong corrections. In the
second one ($\mu>0$) the SUSY-QCD corrections have the same sign
 as the standard QCD ones, so reinforcing the large negative
corrections. In both cases we have fixed $\mu A_{t}<0$, which 
is the overall sign which makes allowance for the low energy data on
radiative $B$ meson decays to be compatible with the existence of a
light charged Higgs below the top quark mass~\cite{bsg}.
This fixes the SUSY-EW corrections~(\ref{eq:dmb}) to be positive. In
Fig.~\ref{fig:tbh}b we present the relative corrections induced by each sector
of the MSSM, in the $\mu<0$ scenario.

The results shown in Fig.~\ref{fig:tbh} can hardly be overemphasized. Whereas
the QCD prediction for the partial decay width states that this is always
significantly smaller than the standard partial decay width, in the $\mu<0$
scenario the charged Higgs partial decay width is equal to the standard one
for $\tb\simeq50$, and it is rapidly increasing. In this
scenario the presence of charged Higgs in top quark decays is significantly
greater than the QCD prediction, and thus the experimental
discovery reach of charged Higgs in top quark decays can be larger than
expected. On the other hand in the $\mu>0$ scenario the
discovery reach is decreased with respect to the standard one. Thus the
excluded region in the $\tb-\mHp$ plane due to the (up to now)
unsuccessful search of charged Higgs bosons in top quark decays
depends drastically on the value of the rest of parameters in the
MSSM~\cite{Guasch:1998jc}.

In Fig.~\ref{fig:tbh}b we see clearly the close-to-linear behaviour of the
leading SUSY contributions~(\ref{eq:dmb}), and we also see that the 
Higgs-boson mediated contributions ($\delta_{EW}$ in the plot)
really play a marginal role. Interestingly enough we see that for $\tb
\simeq35$ the SUSY-QCD corrections cancel the standard QCD ones, and thus,
although the larger corrections are due to the strong interaction sector, the
only radiative corrections that are left are the SUSY-EW ones.

The  SUSY radiative corrections  above $\tb\simeq35$ (at one-loop)  can
easily reach values of
\begin{eqnarray}
&&  \delta_{\rm S-EW}\simeq+30\%\,\,,\,\,\delta_{\rm S-QCD}\simeq+80\%\ \ (\mu
<0,A_{t}>0,M_{SUSY}\simeq100-200\GeV)\,\,,\nonumber\\
&&  \delta_{\rm S-EW}\simeq+20\%\,\,,\,\,\delta_{\rm S-QCD}\simeq-40\%\ \ (\mu
>0,A_{t}<0,M_{SUSY}\simeq500\GeV)\,\,.
\end{eqnarray}
Negative corrections for the SUSY-EW corrections of the same
absolute values are possible provided $\mu A_{t}>0$. We have singled out
different sparticle spectra for the two scenarios in order to avoid total
corrections greater than $100\%$ when they are added to the standard QCD corrections.

\subsection{FCNC top-quark decays}

Flavour Changing Neutral Current (FCNC) decays of the top quark are one-loop
induced processes. They are such  rare events in
the SM~\cite{eilammele}, 
with branching ratios at the level of
 $10^{-10}-10^{-15}$ depending on the particular channel, that its
presence at detectable levels would clearly indicate the presence of new
physics. The question is whether the presence of SUSY particles could enhance
these partial decay widths up
to the visible level. The partial FCNC decay width into a weak vector boson
$\Gamma(t\rightarrow cV)$ ($V=Z\,,\gamma$) undergoes some
enhancement~\cite{tcVweak,deDivitiis:1997sh}, 
however it is still of the order of $10^{-12}-10^{-13}$
in most of the parameter space, thus being far away of the
detection level. The gluon channel ($t\rightarrow cg$) is the
most gifted one in the SM, but it is nevertheless too small to be detectable
($\sim10^{-10}$). This mode, however, has recently been analyzed in 
great detail in the MSSM~\cite{deDivitiis:1997sh,Guasch:1999jp,Guasch:1999ve}
and one finds that its branching ratio can be close to the visible threshold
for the future high luminosity machines such as the LHC and the LC (see
below). Finally, the top quark decaying into neutral Higgs particles
($t\rightarrow ch$, $h=h^{0}$, $H^{0}$, $A^{0}$) has also been shown to
benefit from large enhancements in the MSSM
framework~\cite{Yang:1994rb,Guasch:1999jp,Guasch:1999ve}. In this
respect we recall that in the MSSM the channel in which the lightest Higgs
boson is involved ($t\rightarrow ch^{0}$) is always kinematically
open because $m_{h^{0}}\leq130\,GeV<m_{t}$. Hereafter
we will concentrate on the  two decays, $t\rightarrow cg$ and
$t\rightarrow ch^{0}$, because the overall analysis shows that they
are the most efficient FCNC decays in their respective modalities. Of course
in SUSY models beyond the MSSM, such as models without $R$-parity, there could
be other kind of competing FCNC top quark decays,\footnote{See
e.g.\  Refs.~\cite{FCNCRbreak} 
for some recent works on the
subject.}  but here we shall stick all the time to the MSSM.

FCNC processes can be induced through SUSY-EW charged current interactions.
These proceed through the same mixing matrix elements as in
the SM: the Cabibbo-Kobayashi-Maskawa mixing matrix. But in addition it could
happen that the squark mass matrix squared is not
proportional to the quark-mass-matrix squared. In this case the squark mass
eigenstates  would not coincide with the quark mass states,
and as a consequence  tree-level FCNC would appear in the
quark-squark-gaugino/higgsino interactions. This mixing appears as non-flavour
diagonal mass matrix elements in the squark  mass matrix
squared
\begin{equation}
\label{eq:defdelta}
(M_{LL}^{2})_{ij}=m_{ij}^{2}\equiv\delta_{ij}\,m_{i}\,m_{j}\,\,,\,\,(i\neq
j)\,\,,
\end{equation}
where $i,j$ represent squarks of any generation, and $m_{\{i,j\}}$ is the mass
corresponding to the diagonal entries in the matrix. In the MSSM these kind of
mixing terms  in the Left-chiral sector are naturally
generated through the Renormalization Group evolution of the
soft-SUSY-breaking squark masses down to the EW
scale~\cite{Duncan}. 
This is the reason why we just
singled out the $LL$ mixing component in Eqn.~(\ref{eq:defdelta}). Flavour mass
mixing terms for the corresponding Right-chiral squarks are allowed, but they
do not appear naturally in the GUT frameworks. Moreover its presence 
is not essential because it does not change the order of magnitude of
the results obtained with only flavour-mixing between Left-chiral
squarks~\cite{deDivitiis:1997sh,Guasch:1999jp}. The mixing terms $\delta_{ij}$
are restricted by the low energy data on FCNC
processes~\cite{deltalimits}. 
These limits were computed
using the mass-insertion approximation, so they should be taken as order of
magnitude limits. Recently it has been shown that the full computation of some
FCNC process can give results which differ substantially 
from the mass-insertion approximation ones~\cite{Borzumati:1999qt}.

To assess the size of the FCNC top quark decay rates we use
the fiducial ratio 
\begin{equation}
\label{eq:defbr}
B(\tch)\equiv\frac{\Gamma(t\rightarrow c\,X)}{\Gamma(t\rightarrow b\,W^{+}%
)}\,\,\,,
\end{equation}
for both $X=g,h^{0}$. The typical values of this
ratio lie in the ballpark of $10^{-8}$ for the SUSY-EW contributions
in the regime of large $\tb$\ ($30\lsim\tb\lsim50)$ and for
a SUSY spectrum around $200\GeV$. This is already five orders
of magnitude larger than the corresponding processes in the
SM. The SUSY-QCD contributions, which appear when the $\delta_{23}$ in
Eqn.~(\ref{eq:defdelta}) is non-zero for up-type squarks, are typically around
two orders of magnitude larger, and they exhibit a slow
decoupling as a function of the gluino mass, so even for gluinos as heavy
as $\mg\simeq500\GeV$ the ratio $B(t\rightarrow ch^{0})$ can
reach the level of $10^{-5}$. This large value is due to the
strong nature of the gluino-mediated interactions, but
not less to the fact that present bounds on $\delta_{23}$
are rather poor. Also the various decays are sensitive to both: the higgsino
mass parameter $\mu$ and the soft-SUSY-breaking trilinear coupling $A_{t}$. 

\figtch

In Figs.~\ref{fig:tch}a and b we
present the result of maximizing the ratio~(\ref{eq:defbr}) for the SUSY-EW
and SUSY-QCD contributions respectively~\cite{Guasch:1999jp}. 
These plots have been obtained by performing a full
scan of the MSSM parameter space, in the phenomenologically allowed region,
and for SUSY parameters below $1\TeV$. Needless to say, not all of the maxima
can be simultaneously attained as they are obtained for different values of
the parameters. Perhaps the most noticeable result is that the decay into the
lightest MSSM Higgs boson ($t\rightarrow c\,h^{0}$) is the one that can be
maximally enhanced, reaching values of order $B(t\rightarrow
c\,h^{0})\sim10^{-4}$ that stay fairly stable all over the parameter space, and
in 
particular for almost all the range of allowed Higgs boson
masses in the MSSM.

For the sake of comparison, in Fig.~\ref{fig:tch} we show the
maximized ratio for the competing decay $t\rightarrow cg$ as a function of
the intergenerational mixing parameter between the second and the third
generation, $\delta_{23}$~(\ref{eq:defdelta}). We see that it never really
reaches the critical value $10^{-5},$ which can be considered as the visible
threshold for the next generation of high luminosity
colliders. To assess the discovery reach of the FCNC top quark decays 
for these future accelerators we take as a guide the estimations that
have been made for gauge boson final states~\cite{Frey:1997sg}. Assuming that
all the FCNC decays $t\rightarrow c\,X$ ($X=V,h$) can be treated similarly, we
roughly estimate the following sensitivities for $100\;fb^{-1}$ of integrated
luminosity:
\begin{equation}
\mathrm{\mathbf{LHC:}}B\gsim5\times10^{-5}\,\,;\,\,\mathrm{\mathbf{LC:}}%
B\gsim5\times10^{-4}\,\,;\,\,\mathrm{\mathbf{TEV33:}}B\gsim5\times
10^{-3}\,\,\,.\nonumber\label{sensitiv}%
\end{equation}

Although the LHC seems to be the most sensitive machine to
this kind of physics (due to its highest luminosity) the LC is also very
competitive due to the cleanness of its environment which should allow a much
more efficient isolation of the rare events. The upgraded Tevatron,
unfortunately,  looks not so promising in this respect. 

To better assess the realistic possibilities for detecting the most
serious FCNC top quark decay candidates, $t\rightarrow c\,h$ and
$t\rightarrow c\,g$,  we remark that around the loci of maximal
rates in parameter space the following situation is achieved 
\begin{equation}
5\times10^{-6}\lsim B(t\rightarrow c\,g)_{\max}<B(t\rightarrow c\,h^{0}%
)_{\max}\lsim5\times10^{-4}.\label{maxs}%
\end{equation}
In both types of decays the dominant effects come from SUSY-QCD\@. However, it
should not be undervalued the fact that the maximum electroweak rates for
$t\rightarrow c\,h$ can reach the $10^{-6}$ level. Last but not least, we
stress once again that the largest FCNC rate both from SUSY-QCD and SUSY-EW is
precisely that of the lightest CP-even state ($t\rightarrow c\,h^{0}$), which
is the only Higgs channel that is phase-space available across the whole MSSM
parameter space.

\subsection{Two-body decays into $R$-odd particles}

In principle there exist three possible two-body decays of the top quark into
$R$-odd particles: $t\rightarrow\sbottom\cplus$, $t\rightarrow\stopp\sg$ and
$t\rightarrow\stopp\neut$. These decays were reviewed
in~\cite{borzutdecay}. 
From the latest combined analysis of the four LEP experiments a lower bound
on the squark and chargino masses between $80\GeV$ and $90\GeV$ is
obtained~\cite{LEPSUSY}. The exact bound depends on the
assumptions of the analysis. The chargino channel is then highly disfavoured,
and might be already closed at the end of the present LEP run. The light
gluino window ($\mg\lsim5\GeV$) still exists, but its importance is everyday
more marginal~\cite{Clavelli:1999sm}. Otherwise the direct limits from the
Tevatron $\mg\gsim200\GeV$~\cite{Savoy:1999ui} apply, and the gluino decay
channel is completely ruled out. So we are left with the neutralino decay
channel as the most interesting decay. Note that light top-squarks are a
natural feature in the MSSM due to the large top quark Yukawa coupling, and
the presence of large mixing in the stop sector. A light top-squark (with
small \tb) also leads to an enhancement of $R_{b}$, pushing it closer to the
measured
value~\cite{Rb}. 
Under the conditions of light stop and low \tb\  the branching ratio for the 
top quark decay into a neutralino and a stop can be at the 20\% to 30\% level.
The strong sector one-loop radiative corrections to this partial decay width
have been computed
in~\cite{Djouadi:1996pi,tdecaychar}. 
A key
feature of the radiative corrections in which $R$-odd and $R$-even particles
are both involved in external legs is that it is no longer possible to
separate between standard and SUSY corrections, since neither of these subsets
is finite by itself, so gluon and gluino loop contributions must be added up
in order to obtain a finite result. As a consequence of that there appear
terms which present non-decoupling, namely corrections proportional to
$\alpha_{s}\log(\mg/\msta)$. 

The on-shell renormalization procedure must again deal with non-standard
counterterms. In this case a counterterm for the stop mixing angle, $\delta\theta_t$, is
necessary, even in the case of vanishing angle; this can be fixed by the
condition that $\delta\theta_t$ cancels the one-loop mixing two-point function
between the two stops at $\mstt$~\cite{Djouadi:1996pi}.

\figtstneut
In Fig.~\ref{fig:tstneut}a we see the tree-level prediction for the various
branching ratios $B(t\to\stopp\neut_{\alpha})$ as a function of the lightest
stop mass. We see that large values of this branching ratio can be obtained.
In Fig.~\ref{fig:tstneut}b the radiative QCD corrections to each of the decays
are plotted. Their absolute value lies in the range $2-12\%$, for the small
values of the gluino mass ($\mg\simeq180\GeV$) used in this figure. For larger
gluino masses the corrections are negative, and in the range $\mg=1-5\TeV$
they grow from $-12\%$ to $-22\%$~\cite{Djouadi:1996pi}.

\subsection{Three-particle decays}

In order to have a consistent description of the top quark decay width at the
order $G_{F}\alpha$ it is necessary to account for the three-body decays of
the top quark as well. In Ref.~\cite{Guasch:1997dk} all the
possible three-body decays of the top quark were
investigated.\footnote{See also Ref.~\cite{Belyaev:2000pg} for a recent analysis
  of some of these modes, including $R$-parity violating decays.} 
The aim in that work was to concentrate the
analysis in regions of the parameter space in which the corresponding
two-body decays were phase-space closed. With the latest
bounds on SUSY particle masses~\cite{LEPSUSY} some of 
these decays turn out to be phase-space closed. Here
we briefly review those that still are kinematically allowed.

$t\rightarrow b\tau^{+}\nu_{\tau}$: In the
MSSM there exists, in addition to the gauge boson mediated
channel, also the charged Higgs boson one. Of course, if the charged Higgs is
lighter than the top quark then it would proceed through the two-body decay
$t\rightarrow H^{+}b$ analyzed in section~\ref{sec:tbh}. But if the charged
Higgs is heavier than the top quark, then
the additional contribution to the $\tau^{+}\nu_{\tau}$ final
state from the $H^{+}$ mediated diagram is in the range of 1-3\% 
for $\mHp<200\GeV$ and large $\tb>40$;
 $t\rightarrow bW^{+}h^{0}$: Its decay width can only reach at most
$\Gamma/\Gamma_{SM}^{0}\lsim3\times10^{-4}$. 

Next we just
quote the maximum branching ratios for some decays into $R$-odd particles. 
$t\rightarrow b\neut\cplus$: $\Gamma/\Gamma_{SM}^{0}\lsim0.006$;
$t\rightarrow\sbottom\tau^{+}\tilde{\nu}_{\tau}$: $\Gamma/\Gamma_{SM}^{0}%
\lsim1\%$; $t\rightarrow b\sg\cplus$: We remark that this decay could
only be possible in the light gluino scenario, but even in
this case $\Gamma/\Gamma_{SM}^{0}\lsim4\%$. Let us comment
now a bit on the decay $t\rightarrow b\stau\tilde{\nu}_{\tau}$. This is an
interesting process since, although it contains two $R$-odd particles in the
final state, both of them are sleptons, which therefore evade
the limits from hadron colliders, and being from the third generation they
are expected to be light. Two Feynman diagrams contribute to the amplitude of
this process, one of them involves the exchange of a charged Higgs particle.
Part of the couplings between $H^{+}$ and the $\stau-\sneut_{\tau}$ state
involve the combination $A_{\tau}\tb+\mu$, which can naturally reach large
values. It turns out that if the decay channel $H^{+}\rightarrow\stau
\sneut_{\tau}$ is closed (i.e. $\mHp<m_{\stau}+m_{\sneut_{\tau}}$) the value
if this partial decay width is enhanced. The maximum value of the partial
decay width is obtained for $\mHp$ slightly below the sum of the slepton
masses. With this condition, together with large $\tb\gsim40$, and
$|\mu|\gsim100\GeV$ we find that the values for the corresponding 
branching ratio can easily be of $50\%$~\cite{Guasch:1997dk}. Of
course, this kind of precise alignment may be not the most natural expectation
in the MSSM, but even if the parameters are not ``optimized'' for a maximum
branching ratio, there exist regions (in the large \tb\  scenario) where it can
be relevant, say above the $10\%$ level. One of the possible signals for this
decay channel includes a $\tau^{-}$, i.e.\  a ``wrong sign'' lepton, which
could help in identifying this decay chain. See Ref.~\cite{Guasch:1997dk} for
a complete compilation of possible distinctive signals.

\section{Top-quark pair production in 
       ${e^+e^- \rightarrow  t\bar{t}}$}
\label{sec:prod}

Besides the possibility of direct production of SUSY particles at
sufficiently high energies, accurate investigations of   
the production of standard fermion pairs
in $e^+e^-$ annihilation offers the indirect search for
virtual SUSY particles through quantum effects in terms of loop 
corrections.
In particular the top quark, owing to its large mass, is an ideal
probe of all those virtual effects that grow with the fermion mass
scale, such as Yukawa interactions and CP violation.
The MSSM with complex parameters induces CP-violating observables
already at the one-loop level; they are described below after 
a brief discussion of the generic supersymmetric loop contributions 
to the CP-conserving cross section for $t\bar{t}$ production.

\subsection{Loop effects in the CP-conserving MSSM}

The process of $t\bar{t}$ production
is in lowest order described by photon and $Z$-boson exchange.
The cross section can be written as follows
\beq
\label{xsecborn}
    \sigma^{(0)}(e^+e^- \rightarrow t\bar{t}) \, 
      = \, \frac{\beta}{4\pi s}\, \left[
    \frac{3-\beta^2}{2} \, \sigma_V(s) + \beta^2 \, \sigma_A(s) \right]   
\eeq
with 
\beq 
 s = (p_{e^-}+p_{e^+})^2, \quad
 \beta = \sqrt{1-\frac{4m_t^2}{s}} \, , 
\eeq
and 
\beq
 \sigma_V & = & 
   Q_e^2 Q_t^2\, e(s)^4 \, + \, 
           2 v_e v_t Q_e Q_t\, e(s)^2\,  
            \chi(s)\, + \,
           (v_e^2+a_e^2) v_t^2 \, \chi(s)^2 , \nonumber \\
 \sigma_A  &= &  (v_e^2+a_e^2)\, a_t^2 \, \chi(s)^2 \, .
\eeq
This expression is an effective Born approximation, which incorporates
the effective (running) electromagnetic charge containing the 
photon vacuum polarization (real part) 
\beq
 e(s)^2 = \frac{4\pi\alpha}{1-\Delta\alpha(s)}\,  ,
\eeq
the $Z$ propagator, together with the overall normalization factor of the
neutral-current couplings in terms of the Fermi constant $G_\mu$,  
\beq
 \chi(s) = (G_\mu M_Z^2 \sqrt{2})^2\, 
                 \frac{s}{s-M_Z^2 }  \, ,
\eeq 
and
the vector and axial-vector coupling constants for $f=e, t$,    
\beq 
v_f = I_3^f -2 Q_f \sin^2\theta_W , \quad
a_f = I_3^f \, .
\eeq
The complete electroweak one-loop corrections were calculated  
for the case of the Standard Model~\cite{hollik91,hollik91a}, 
including also the hard-photon QED corrections~\cite{leike}. 
In a more recent  study~\cite{hs}, a
complete one-loop calculation was performed for the MSSM
electroweak corrections and the non-standard part of the QCD corrections
(SUSY-QCD corrections) to $e^+e^- \to t \bar{t}$.
The virtual Higgs contributions 
have also been derived in a general 2-Higgs-doublet model~\cite{arnd}. 
For a large mass of the $A^0$ boson
in the MSSM Higgs sector, the heavy particles $H^0, A^0, H^\pm$
decouple and
the light $h^0$ scalar behaves like the standard Higgs boson.
In this so-called decoupling limit, the only non-standard virtual
SUSY effects arise from the genuine supersymmetric particles.  
 
The one-loop contribution
to the $S$-matrix element contains the 
$\gamma$ and $Z$ self-energies, the $\gamma$ and $Z$ vertex corrections
together with the external wave function renormalization, and the box 
diagrams.
Since Higgs-boson couplings to the initial state $e^+,e^-$ 
are negligible, we only have to consider
the standard box graphs with $Z$ and $W^\pm$ exchange and the SUSY box
graphs with neutralino and chargino exchange.

The complete set of vertex corrections comprises the QED corrections 
with virtual photons and the QCD corrections with virtual gluons.
They need real photon and gluon bremsstrahlung for
a infrared-finite result. The gauge-invariant subclasses of 
``standard QED'' and ``standard QCD'' corrections are identical to
those in the Standard Model and are available in the literature~\cite{hollik91,leike,Zerwas}. 
In the meantime, also a combined treatment
of the QCD and the electroweak radiative corrections in the Standard Model
has been proposed~\cite{hahn} in order to account for the dominant
terms from both sources. In our context, we are interested in 
deviations from the Standard Model induced by non-standard virtual particles.
We therefore concentrate our discussion on the
set of model-dependent and IR-finite virtual corrections, 
without the diagrams involving 
virtual photons and gluons.
The supersymmetric part of the QCD corrections, arising from virtual 
gluinos, is included  as part of the final-state 
vertex correction.

\begin{figure}
\begin{center}
\includegraphics[width=10cm,clip]{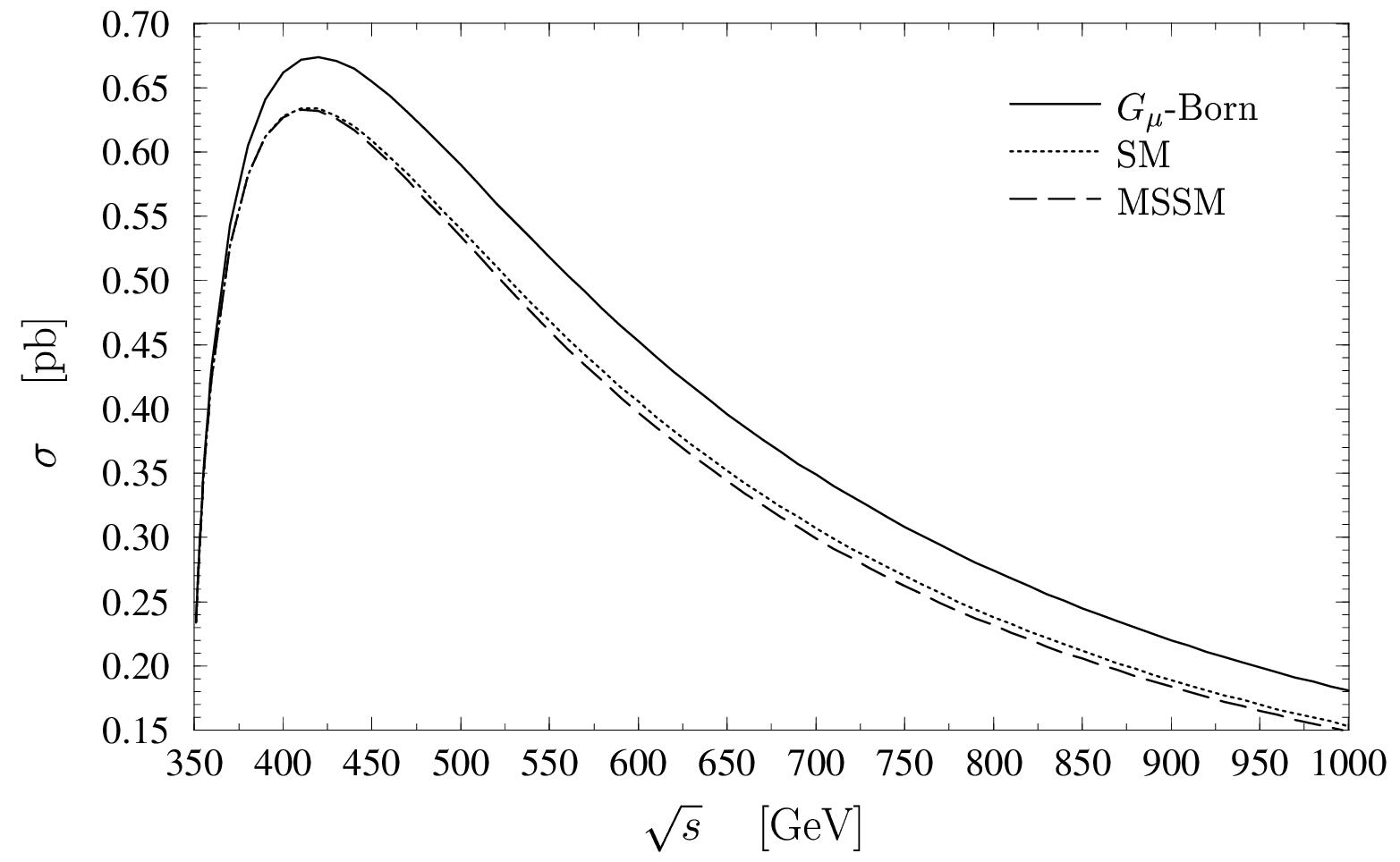}
\end{center}
\vspace*{-0.5cm}
\caption{\sf The cross section for $e^+e^- \rightarrow t \bar{t}$,
as a function of the energy.
$\sqrt{s} = 500\, \gev$, $M_A = 500\, \gev$, $m_{\tilde{t}_1} = 100\,\gev$,
$m_{\chi_1^+} = 100\, \gev$,  $A= 500\, \gev$, $\mu = -150\, \gev$,
$\tan\beta = 40$.}
\label{eetts}
\end{figure}

\begin{figure}
\begin{center}                                    
\includegraphics[width=10cm,clip]{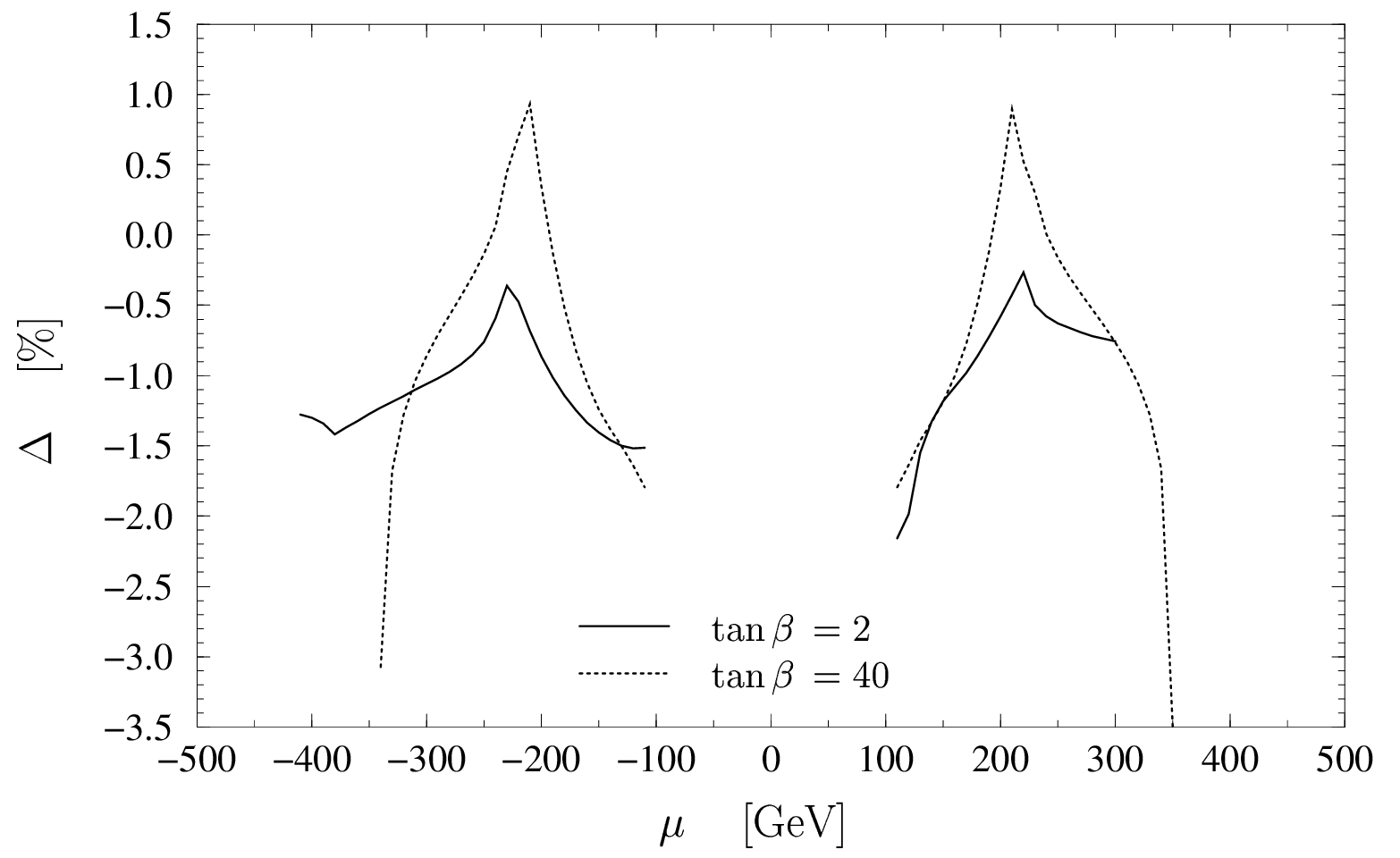} 
\end{center}
\vspace*{-0.5cm}
\caption{\sf Difference between MSSM and SM, relative to the Born cross 
section, for a light stop and chargino. 
$\sqrt{s} = 500\, \gev$, $M_A = 500\, \gev$, $m_{\tilde{t}_1} = 100\, \gev$,
$m_{\chi_1^+} = 100\, \gev$, $A= 500\, \gev$. }
\label{Deltalight}
\end{figure}

\begin{figure}
\begin{center}
\includegraphics[width=10cm,clip]{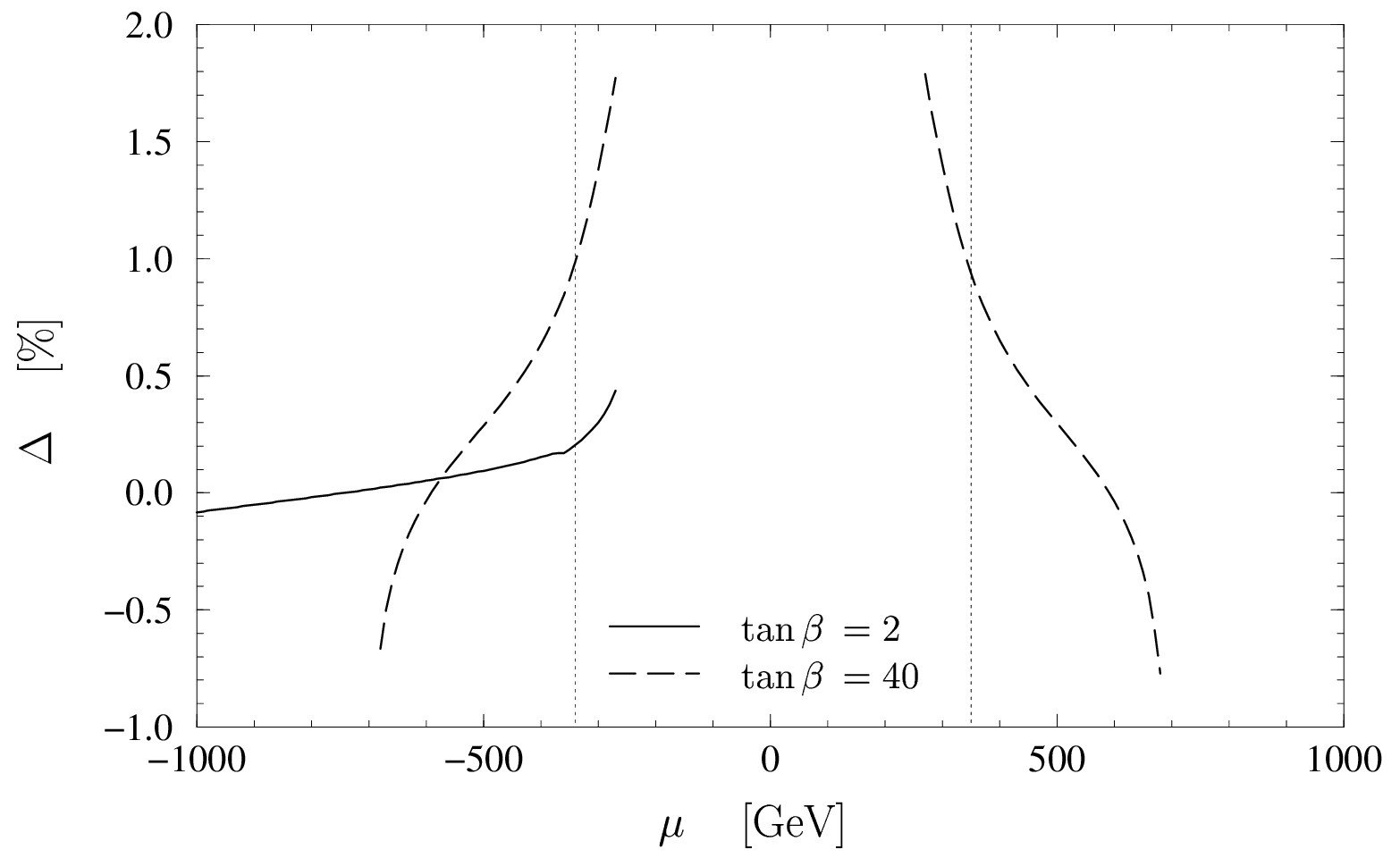} 
\end{center}
\vspace*{-0.5cm}
\caption{\sf Difference between MSSM and SM, relative to the Born cross 
section for heavy SUSY particles.
$\sqrt{s} = 500\, \gev$, $M_A = 500\, \gev$, $m_{\tilde{t}_1} = 260\, \gev$,
$m_{\chi_1^+} = 260\, \gev$, $A= 500\, \gev$. 
The range consistent with the
constraints is between the vertical lines for the large $\tan\beta$
case only; the solid line is not constrained. }
\label{Deltaheavy}
\end{figure}

The effect of these IR-finite loop-corrections is illustrated in 
Figure~\ref{eetts}, where the energy dependence of 
lowest-order cross section~(\ref{xsecborn}) 
is shown together with the Standard Model (SM)  and the MSSM 
one-loop prediction. The differences between the SM and the 
MSSM depend on the values of the SUSY-breaking parameters  and
the Higgsino  mass parameter~$\mu$. In case of a direct detection of
SUSY particles, these parameters can, at least partially, 
be measured and the top production cross section can provide
an independent test of the MSSM. In a first phase of a Linear Collider
operation, the mass of eventually light SUSY particles, such as 
a scalar top $\tilde{t}_1$ or a chargino $\chi_1^\pm$ may be determined
sufficiently well, while the heavier particles 
may appear as less accessible. A precise measurement of the $t\bar{t}$ cross
section is therefore useful to probe the heavier part of the model
through the quantum contributions. To make this more quantitative,
we assume a scenario with  a relatively 
light stop and chargino, and analyze the dependence of the loop-contributions
on the residual parameters of the MSSM, which determine the 
heavier part of the mass spectrum.
In order to exhibit the deviations from the SM, we introduce the
quantity
\beq
  \Delta = \frac{\sigma^{\rm MSSM}(s) - \sigma^{\rm SM}(s)}
                {\sigma^{(0)}(s) } \, ,
\eeq
which gives the difference between the models normalized to the
Born cross section~(\ref{xsecborn}).
The results are displayed in Figure~\ref{Deltalight}.
For simplicity, we have assumed a common mass scale $M_S$ 
and a common  $A$ parameter for the diagonal and non-diagonal 
entries in the sfermion mass-square matrices
of all generations (as well as for $L$ and $R$ chiralities). 
The free parameters in Figure~\ref{Deltalight} are varied in
accordance with a set of constraints which ensure consistency with 
the present electroweak precision data and bounds from direct Higgs searches.
The effects are in the per-cent range, and show quite some sensitivity
to the model parameters. Large values for $\tan\beta$ induce bigger
loop effects, owing to the enhanced $b$ quark/squark Yukawa couplings.

As another possible scenario, we consider the situation that  SUSY
particles are not directly detected at a 500 GeV collider. 
The corresponding virtual SUSY effects, compatible with these bounds 
and with the constraints from the electroweak precision data, 
are shown in Figure~\ref{Deltaheavy}.  
At least for the large-$\tan\beta$ regime the loop contributions
are at the level of 1--2 per cent.

\subsection{CP violation in the MSSM}

\subsubsection{The MSSM with complex couplings}

The following parameters of the MSSM with preserved $R$-parity may take 
complex values: the Yukawa couplings, the $\mu$ parameter and the soft-breaking 
parameters $m_{12}^2$ (in the Higgs potential), $M_1$, $M_2$, $M_3$ 
(gaugino mass terms) and $A$ (trilinear terms).
But not all of the complex phases introduced are physical since several of 
them can be absorbed by redefinitions of the fields. 

We assume the GUT relation between the $M_i$, so they have one common phase. 
The remaining phases are chosen to be those of $\mu$, the $A$ 
parameters and, for three generations, one phase for all the Yukawa couplings,
the $\delta_{\rm CKM}$.
Analogous to the Standard Model case, the Yukawa couplings can be
changed by redefinitions of the quark superfields in such a way that
there remains only one phase for three generations.

Furthermore, the ${\cal L}_{\rm MSSM}$ is invariant under two U(1) 
transformations, the Peccei-Quinn symmetry and the $R$-symmetry, that do not 
only transform the fields but also the parameters of the MSSM~\cite{relax}. 
With them one can easily show~\cite{hirs3} that the physical predictions only 
depend on the absolute values of the parameters and the phases
\beq
\phi_A\equiv {\rm arg}(A M_i^*), \quad \phi_B\equiv {\rm arg}(\mu A
m_{12}^{2*}).
\eeq

It is then possible to choose a set where only the $\mu$ and the $A$ 
parameters are complex. We take $\varphi_\mu={\rm arg}(\mu)$ and the phase 
of $A$ is traded for the phase $\varphi_{\tilde f}$ of the off-diagonal term 
in the corresponding sfermion mixing matrix: 
$m^f_{LR}\equiv A_f-\mu^*\{\cot,\tan\}\beta$, for \{up, down\}-type fermions, 
respectively. Relaxing universality for the soft-breaking terms, 
every $A_f$ has a different phase. 

In comparison with the MSSM with real couplings, the new physical phases
affect the mass spectrum of charginos and neutralinos, modify the
tree level couplings and notably produce CP-violating effects at the one-loop 
level that we analyze in the following.

\subsubsection{Electric dipole form factors}

The most general Lorentz structure for the vertex $Vff$ in the momentum space is
\beq
\Gamma^{Vff}_\mu & = & {\rm i}  \Big[
\gmu \left(f_{\rm V} - f_{\rm A} \gfi \right) \ + \ 
(q-\bar{q})_\mu \left(f_{\rm M} + {\rm i}f_{\rm E} \gfi \right) + \ p_\mu 
\left({\rm i}f_{\rm S} + f_{\rm P} \gfi \right) \nn \\ 
& & \qquad  +
(q - \bar{q})^\nu \smunu 
\left( f_{\rm TS} + {\rm i}f_{\rm TP} \gfi \right) \ + \ 
p^\nu \smunu \left({\rm i}f_{\rm TM} + f_{\rm TE} \gfi 
\right) \Big],
\label{efflagm}
\eeq
where $q$ and $\bar{q}$ are the outgoing momenta of the fermions and 
$p=(q+\bar{q})$ is the total incoming momentum of the neutral vector boson $V$. 
The form factors $f_i$ are functions of kinematical invariants and
can be complex in general. Their real parts account for dispersive effects 
(CPT-even) whereas their imaginary parts are related to absorptive 
contributions. For on-shell fermions, making use of the Gordon identities,
one can eliminate $f_{\rm TM}$, $f_{\rm TE}$, $f_{\rm TS}$ and $f_{\rm TP}$.
The number of relevant form factors can be further reduced when 
$V$ is on shell, since then the condition $p_\mu \eps^\mu=0$ 
cancels the contributions coming from $f_{\rm S}$ and $f_{\rm P}$.
The form factors $f_{\rm V}$ and $f_{\rm A}$ are connected to the chirality 
conserving, CP-even sector. The anomalous magnetic and electric dipole
form factors are defined, respectively, by
\beq
\mbox{AMDFF}&\equiv& a^{V}_f(s)\equiv -2m_f f_{\rm M}(s),\\
 \mbox{EDFF}&\equiv& d^{V}_f(s)\equiv e f_{\rm E}(s).
\eeq 
They are related to chirality-flipping operators of dimension
larger than four. In a renormalizable theory they can receive contributions 
exclusively by quantum corrections. Unlike the MDFFs, the EDFFs are connected 
to the CP-odd sector and therefore constitute a source of CP violation.
The {\em electromagnetic} and {\em weak dipole moments} are physical and gauge 
invariant quantities corresponding to the values of the dipole form factors 
at $s=(q+\bar q)^2=M^2_V$ for $V=\gamma,Z$, respectively.

All the possible one-loop contributions to the $a^V_f(s)$ and $d^V_f(s)$ 
form factors can be classified in terms of the six classes of triangle 
diagrams. 
The vertices involved are labeled by generic couplings according to a 
general interaction Lagrangian.
Every class of diagrams has been calculated analytically and expressed, 
in the 't Hooft-Feynman gauge, in terms of masses, generic (complex) couplings 
and one-loop 3-point integrals in Refs.~\cite{hirs1,hirs2}.

We concentrate here on the CP-violating electric dipoles of the top quark. 
Unlike the SM, where the EDFFs arise first at three loops~\cite{smedm},
in the MSSM they receive contributions already at the one-loop
level~\cite{hirs3,hirs2,vienna}. On the other hand, since the top quark is too
heavy  
to be pair-produced by $Z$ decays, and has a too short life to study its
electromagnetic static properties, the electric $d^\gamma_t(s)$ and 
weak-electric $d^Z_t(s)$ form factors, $s>4m^2_t$, are {\em not 
physical}. In fact, they do not contribute alone to CP-violating observables,
as it will be show in next Section. Nevertheless they happen to be
gauge invariant quantities at one loop. 

\begin{figure}[ht]
\begin{center}
\epsfig{file=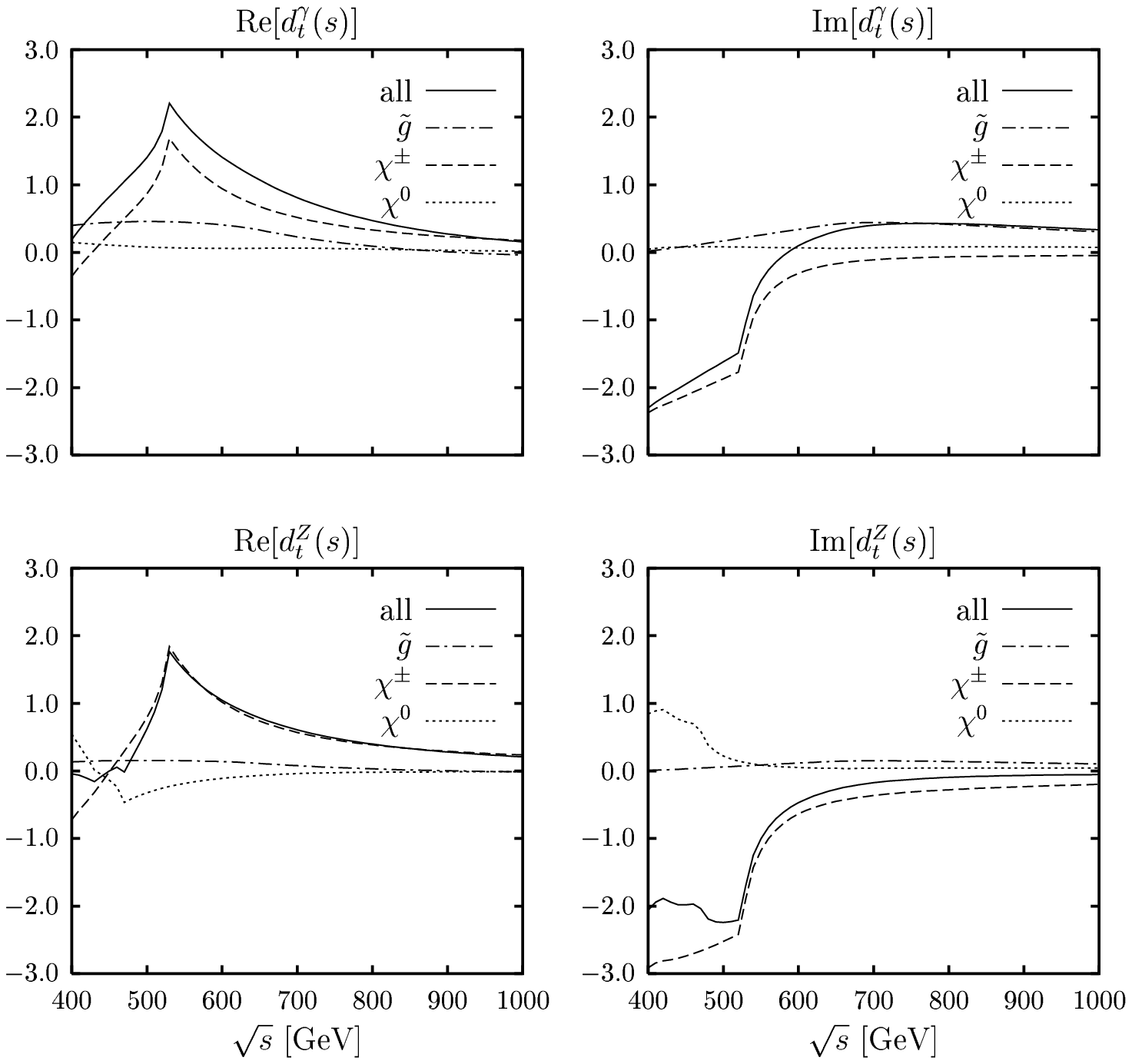,width=0.8\linewidth}
\end{center}
\caption{\sf The contributions to the $t$ (W)EDFF/$10^{-3}\mu_t$ for 
Set \#1 of MSSM parameters (\ref{refset1}).
\label{fig:4.3}}
\end{figure}

The size of the electric dipole form factors of the top quark is illustrated 
in Fig.~\ref{fig:4.3} where the different contributions from charginos,
neutralinos and gluinos are also displayed. Neutralinos and gluinos appear
with stops in the loops, and charginos with sbottoms. The Higgs sector does 
not play any role. The results depend on a number
of MSSM parameters: $\tan\beta$, $M_2$, $|\mu|$, $m^t_{LR}$, $m^b_{LR}$
and $m_{\tilde q}$ (the common squark mass parameter); and three physical
CP-violating phases, $\varphi_{\mu}$, $\varphi_{\tilde t}$ 
and $\varphi_{\tilde b}$. They are expressed in $t$ magnetons $\mu_t\equiv 
e/2m_t=5.64\times 10^{-17}\ e$~cm. The values shown in the plots are for
the set of inputs:
\beq
& &\tan\beta=1.6 \nn \\
& &M_2=|\mu|=m_{\tilde{q}}=|m^t_{LR}|=|m^b_{LR}|=200 \mbox{ GeV} \nn \\
\mbox{Reference Set }\#1:
& &\varphi_\mu=-\varphi_{\tilde t}=-\varphi_{\tilde b}=\pi/2.
\label{refset1}
\eeq
This set of parameters is plausible and enhances the electric dipole form 
factors in the vicinity of $\sqrt{s}=500$~GeV, due to threshold effects 
(spikes in Fig.~\ref{fig:4.3}). The phases are chosen maximal and with such a 
sign that the typically larger components (charginos and gluinos) add up 
constructively. The low $\tan\beta$ scenario has been chosen because both
chargino and gluino contributions are then larger. Of course, the electric
dipoles vanish for zero phases. Due to the decoupling of the supersymmetric
sector, the dipoles tend to vanish for increasing values of the mass parameters. 
A detailed study can be found in~\cite{hirs3}.

\subsubsection{CP-odd observables}

\begin{figure}[ht]
\begin{center}
\epsfig{file=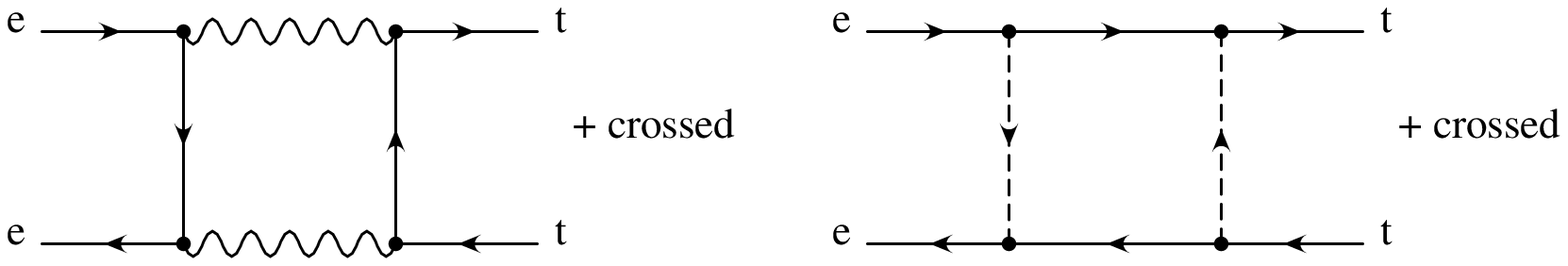,width=12cm}
\end{center}
\caption{\sf Relevant generic box graphs for the MSSM at one loop. The first 
             class involves only SM particles and does not contribute to CP 
             violation.
\label{boxgraphs}}
\end{figure}

Not only vertex- but also box-diagrams correct the tree level process to one 
loop. Accordingly, any CP-odd observable will depend on the 
CP-violating effects due to vertex corrections included in the electric and 
weak-electric dipole form factors as well as on CP-violating box 
contributions. In the former there appear either sneutrinos, neutralinos and
stops or selectrons, charginos and sbottoms.

To probe CP violation, consider the pair-production of polarized $t$ quarks 
$e^+({\bf p_+})+e^-({\bf p_-})\to t({\bf k_+},\sone)+\bar{t} ({\bf k_-},\stwo)$.
The decay channels labeled by $a$ and $c$ act as spin analyzers 
in $t + \bar t\to a(\qplus) + \bar c(\qminus)+X$.
The momenta and polarization vectors in the overall c.m.s. transform under
CP and CPT as follows:\footnote{T means reflection of spins and momenta.}
\beq
 \ba{rl}
 \mbox{CP}:& \ppmn\to -\pmpn=\ppmn \\
           & \kpmn\to -\kmpn=\kpmn \\
           & \qpmn\to -\qmpn \\
           & \sone\leftrightarrow \stwo \\
 \ea
&\hspace{1cm}&
 \ba{rl}     
\mbox{CPT}:& \ppmn\to \pmpn=-\ppmn \\
           & \kpmn\to \kmpn=-\kpmn \\ 
           & \qpmn\to \qmpn \\
           & \sone\leftrightarrow -\stwo 
 \ea
\eeq
From the unit momentum of one of the $t$ quarks in the c.m.s.
(say $\kplus$) and their polarizations 
($\sone$, $\stwo$) a basis of linearly independent CP-odd {\em spin
observables} can be constructed~\cite{bernre91b}. The spin observables are 
related to more realistic (directly measurable) {\em momentum observables}
based on the momenta of the top decay products~\cite{bernre89b}. 
The polarizations can be analyzed through the angular correlation of the weak 
decay products, both in the nonleptonic and in the semileptonic channels:
\beq 
&&t(\kplusn)\to b({\bf q_b}) X_{\rm had}({\bf q_{X}}), \\  
&&t(\kplusn)\to b({\bf q_b})\ell^+(\qplus)\nu_\ell\quad (\ell=e,\mu,\tau)
\eeq
and the charged conjugated ones.

\subsubsection{Spin observables}

\begin{table}[ht]
\begin{center}
\begin{tabular}{|c|l|c|c|c|r|r|r|r|}
\hline
$i$ &CPT & ${\cal O}_i$ & {\bf a}&{\bf b} & 
\multicolumn{2}{|c|}{Set $\#1$} &
\multicolumn{2}{|c|}{Set $\#2$} \\
\hline \hline
1 & even&  $(\soner-\stwor)_y$ & T$\uparrow$&T$\downarrow$
        & 1.216 &$-0.231$& 0.207 &$-1.394$      \\
2 & even&  $(\soner\times\stwor)_x$& T$\uparrow$&L$\uparrow$    
        &$-0.755$&$-0.489$&$-0.053$&$0.318$       \\
3 & even&  $(\soner\times\stwor)_z$& N$\uparrow$&T$\uparrow$    
        & 1.184 &$0.625$&0.090  &$-0.598$       \\
4 & odd &  $(\soner-\stwor)_x$     & N$\uparrow$&N$\downarrow$
        &$-1.230$&$-1.421$&$-1.888$&$-2.217$ \\
5 & odd &  $(\soner-\stwor)_z$     & L$\uparrow$&L$\downarrow$
        & 2.550 &$2.739$ & 3.823 &4.216   \\
6 & odd &  $(\soner\times\stwor)_y$& L$\uparrow$&T$\downarrow$
        &$-1.683$&$-1.751$&$-2.050$&$-2.216$ \\
\hline
\end{tabular}
\end{center}
\caption{\sf Ratio $r\equiv\langle{\cal O}\rangle/\sqrt{\langle{\cal O}^2
\rangle}$ [in $10^{-3}$ units] of the integrated spin observables at $\sqrt{s}=
 500$ GeV for the Set \#1 (\ref{refset1}) and Set \#2 (\ref{refset2}) of
 MSSM parameters. The left column excludes the box corrections and the right 
 one comes from the complete one-loop cross section for $e^+e^-\to t\bar t$.\label{tab1}}
\end{table}

\begin{figure}
\begin{center}
Reference Set $\#1$ \\ ~ \\
\epsfig{file=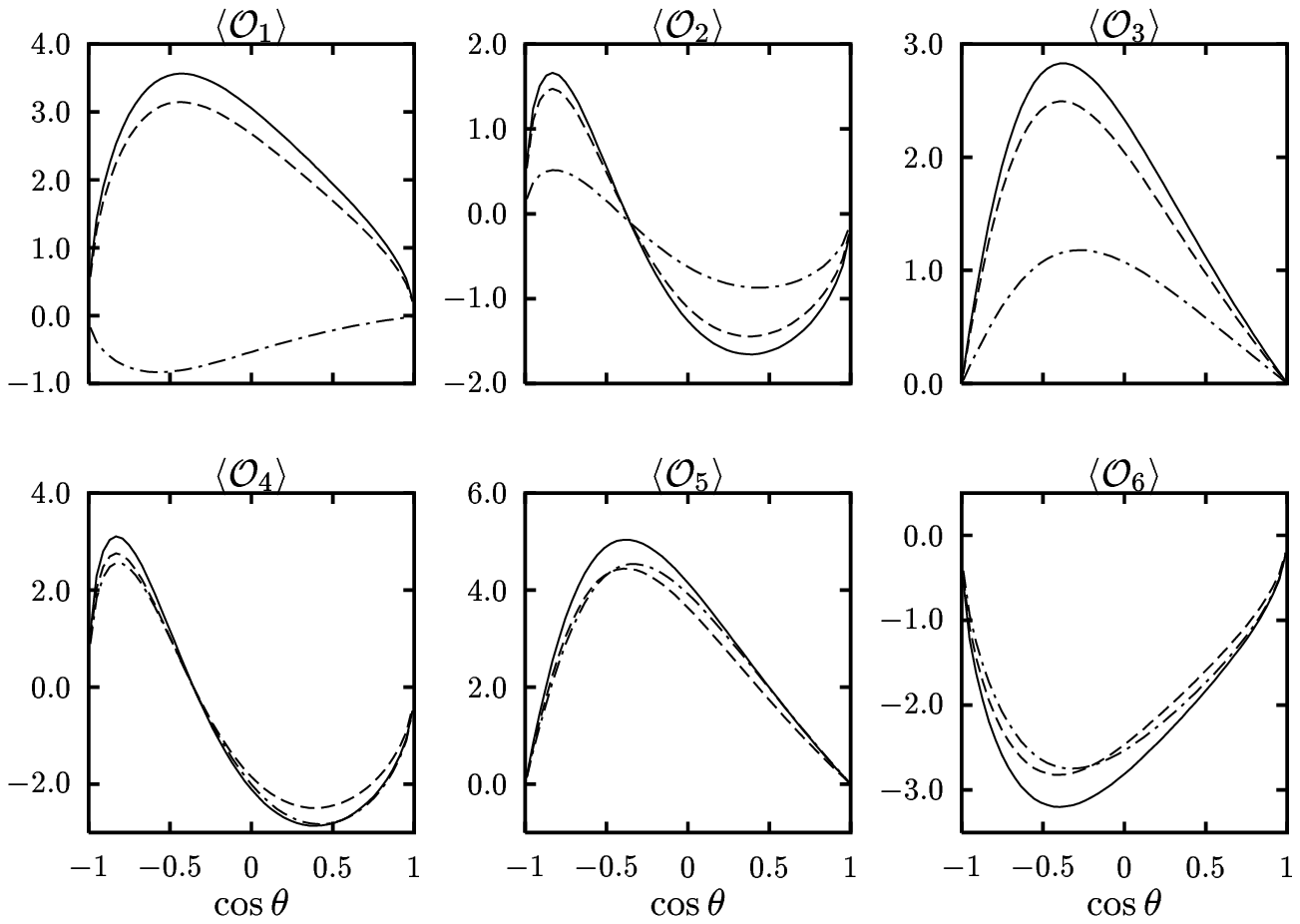,width=0.65\linewidth}
\\ 
Reference Set $\#2$ \\ ~ \\
\epsfig{file=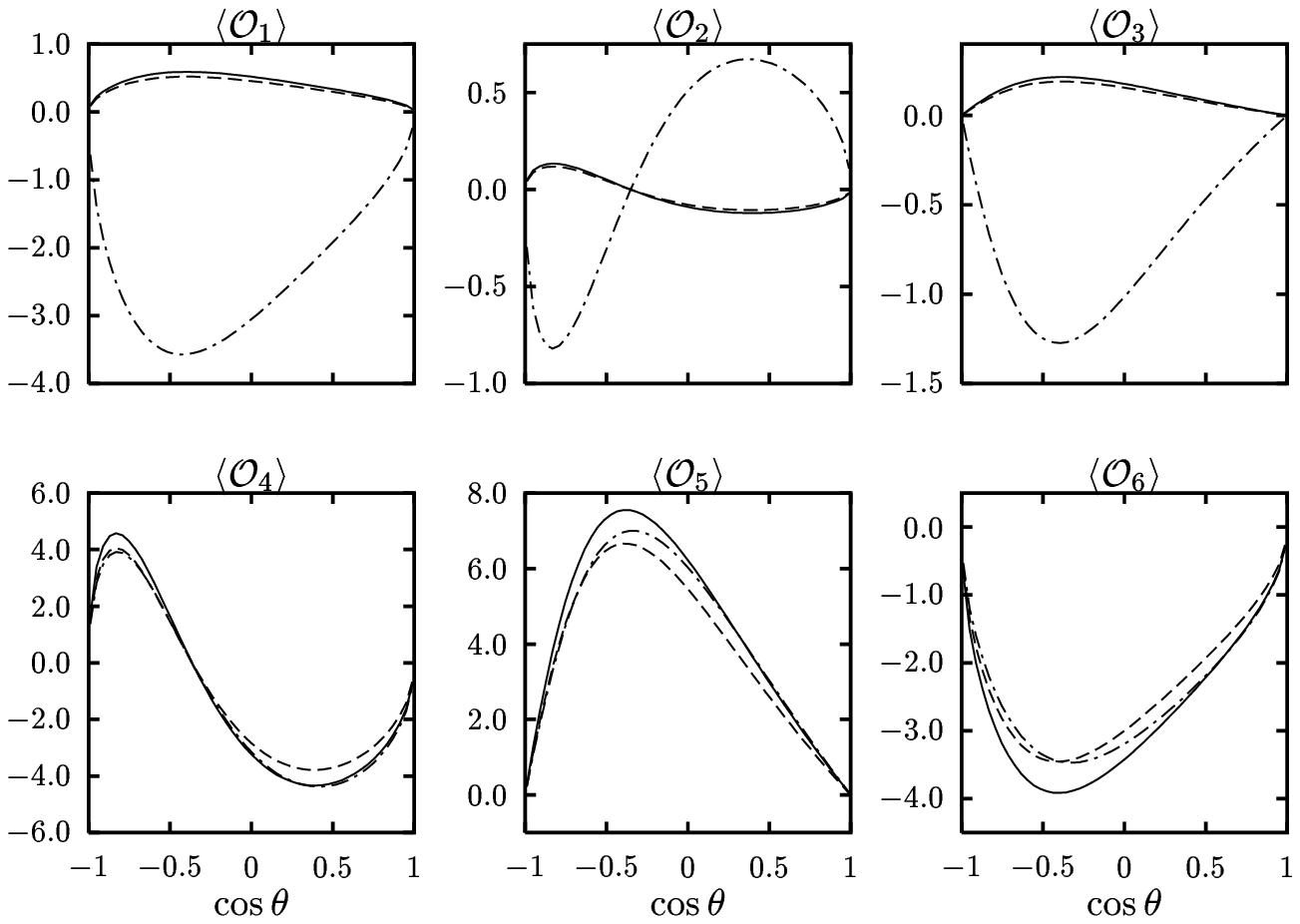,width=0.65\linewidth}
\end{center}
\caption{\sf Expectation value of the spin observables [in $10^{-3}$ units] for
the two reference sets of MSSM parameters, assuming for the cross section: 
  (i) tree level plus contribution from (W)EDFFs only (solid line);  
 (ii) one loop including all the vertex corrections and the self-energies
      (dashed line);
(iii) complete one loop (dot-dashed line).
\label{fig2}}
\end{figure}

A list of CP-odd spin observables classified according to their CPT properties
is shown in Table~\ref{tab1}.
Their expectation values as a function of $s$ and the scattering angle of the
$t$ quark in the overall c.m. frame are given by
\beq
\langle{\cal O}\rangle_{\mathbf{ab}} &=& \frac{1}{2\dd\sigma}\left[
 \sum_{\soner,\stwor=\pm \mathbf{a},\pm \mathbf{b}}
+\sum_{\soner,\stwor=\pm \mathbf{b},\pm \mathbf{a}}\right]
 \dd\sigma(\soner,\stwor)\ {\cal O},
\label{obsaverage} \\
\dd\sigma&=&\sum_{\pm\soner,\pm\stwor}\dd\sigma(\soner,\stwor).
\label{spinaverage}
\eeq
They are displayed in Fig.~\ref{fig2}, for two sets of MSSM parameters,
Set \#1 of (\ref{refset1}) and Set \#2 with same moduli but different phases:
\beq
\mbox{Reference Set }\#2:& &\varphi_\mu=\varphi_{\tilde t}=\varphi_{\tilde b}
=\pi/2.
\label{refset2}
\eeq
The directions of polarization of $t$ and $\bar t$ ({\bf a} and {\bf b}) are 
taken normal to the scattering plane (N), transversal (T) or 
longitudinal (L). They are taken either parallel ($\uparrow$) or antiparallel 
($\downarrow$) to the axes defined by $\hat{z}={\bf k_+}$, $\hat{y}={\bf k_+}
\times {\bf p_+}/|{\bf k_+}\times{\bf p_+}|$ and $\hat{x}=\hat{y}\times\hat{z}$.
The spin vectors $\soner$, $\stwor$ are defined in the $t$, $\bar t$ rest
frames, respectively.
The statistical significance (number of standard deviations) of the 
CP-violation signal is given by 
\beq
N_{SD}=|r|\sqrt{N},\quad r\equiv\langle{\cal O}\rangle/\sqrt{\langle{\cal O}^2
\rangle}
\eeq 
where $N$ is the number of observed events.

We compare the result when only the self energies and vertex corrections
are included (left column) with the complete one-loop calculation (right 
column). The shape of the observables as a function of the $t$
polar angle is also quite different when the CP violating box contributions
are taken into account (Fig.~\ref{fig2}). This illustrates that the {\em dipole 
form factors} of the $t$ quark are {\em not sufficient to parameterize 
observable CP-violating effects} and the predictions can be wrong by far.

\subsubsection{Momentum observables}

Consider now the decay channels labeled by $a$ and $c$ acting as spin analyzers 
in $t + \bar t\to a(\qplus) + \bar c(\qminus)+X$. 
We ignore here possible CP violation in the top-quark decays, that may
occur due to supersymmetric one-loop corrections~\cite{topdecaycp}.
The expectation value of a realistic CP-odd observable is given by the average
over the phase space of the final state particles,
\beq
\langle{\cal O}\rangle_{ac} = \frac{1}{2}\left[ 
\langle{\cal O}\rangle_{a\bar c} + \langle{\cal O}\rangle_{c\bar a} \right] =
\frac{1}{2\sigma_{ac}}\int
\left[ \dd\sigma_{a\bar c} + \dd\sigma_{c\bar a} \right]\ {\cal O},
\eeq
where both the process ($a\bar c$) and its CP conjugate ($c\bar a$) are 
included and 
\beq
\sigma_{ac}=\int\dd\sigma_{a\bar c}=\int\dd\sigma_{c\bar a} ,
\eeq
The differential cross section for $t$-pair production and decay is
evaluated for every channel using the narrow width approximation.
As $m_t>M_W+m_b$, the $t$ quark decays proceed
predominantly through $Wb$. Within the SM the angular distribution of the
charged lepton is a much better spin analyzer of the $t$ quark than that
of the $b$ quark or the $W$ boson arising from semileptonic or nonleptonic 
$t$ decays~\cite{analtopspin}.
The dimensionless observables are easier to measure, for instance
the scalar CP-odd observables~\cite{bernre91b}:
\beq
\hat{A}_1&\equiv&\pplus\cdot\dfrac{\qplusn\times\qminusn}
                                  {|\qplusn\times\qminusn|} 
 \hspace{1cm} \mbox{[CPT-even]} 
\label{A1}\\
\hat{A}_2&\equiv&\pplus\cdot(\qplusn+\qminusn)
 \hspace{1cm} \mbox{[CPT-odd]}
\label{A2}
\eeq 
or the CP-odd traceless tensors~\cite{bernre91b}:
\beq
\hat{T}_{ij}&\equiv&(\qplusn-\qminusn)_i
       \frac{(\qplusn\times\qminusn)_j}{|\qplusn\times\qminusn|} + 
       (i\leftrightarrow j) \hspace{1cm} \mbox{[CPT-even]} 
\label{T33}\\
\hat{Q}_{ij}&\equiv&(\qplusn+\qminusn)_i(\qplusn-\qminusn)_j +
       (i\leftrightarrow j) \hspace{1cm} \mbox{[CPT-odd]}
\label{Q33}
\eeq
The reconstruction of the $t$ frame is not necessary
for the momentum observables. The observables
$\hat{A}_2$ and $\hat{Q}_{ij}$ do not involve angular correlations as
they could be measured considering separate samples of events in the
reactions $e^+e^-\to aX$ and $e^+e^-\to \bar{a}X$. Nevertheless it is 
convenient to treat them in an event-by-event basis~\cite{bernre91b}.

In Table~\ref{tab9} the ratio $r$ is shown for three different decay channels 
at $\sqrt{s}=500$ GeV. The CP-odd observables involve
the momenta of the decay products analyzing $t$ and $\bar t$ polarizations 
in the laboratory frame. The leptonic
decay channels are the best $t$ spin analyzers but the number of leptonic
events is also smaller. The Reference Set $\#2$ has 
been chosen. As expected, the dipole contributions (left columns) to the CPT 
even observables are very small for this choice of MSSM parameters 
but the {\em actual} expectation values (right columns) are larger.

\begin{table}[ht]
\begin{center}
\begin{tabular}{|l|c|r|r|r|r|r|r|}
\hline
CPT     & ${\cal O}$ & \multicolumn{2}{|c|}{$b-b$} 
        & \multicolumn{2}{|c|}{$\ell-b$} 
        & \multicolumn{2}{|c|}{$\ell-\ell$} \\ 
\hline \hline
even& $\hat{A}_1$   &
$-0.036$        &0.242          &
0.030           &$-0.202$       &       
0.068   &$-0.467$\\
odd & $\hat{A}_2$       &  
0.270           &0.304  &
$-0.180$        &$-0.204$&      
$-0.501$        &$-0.812$\\
even& $\hat{T}_{33}$&
$-0.006$        &$0.042$&
0.021   &$-0.140$&      
$-0.037$        &$0.248$\\
odd & $\hat{Q}_{33}$&
0.486           &0.542  & 
$-0.335$        &$-0.374$&      
$-1.274$        &$-1.420$\\
\hline
\end{tabular}
\end{center}
\caption{\sf Ratio $r$ [in $10^{-3}$ units] of the momentum observables 
at $\sqrt{s}=500$ GeV for three different channels: $t + \bar t\to a(\qplus) + 
\bar c(\qminus)+X$, given the Set $\#2$ of MSSM parameters. 
The left column includes only the $t$ (W)EDFF corrections  
and the right one comes from the complete one-loop cross section for 
$e^+e^-\to t\bar t$. \label{tab9}}
\end{table} 

\subsubsection{Polarized beams}

The previous results were obtained for unpolarized electron and positron beams.
Let $P_\pm$ be the degree of longitudinal polarization of the initial
$e^\pm$. The differential cross section reads now
\beq
\dd\sigma=\frac{1}{4}\left[(1+{\rm P}_+)(1+{\rm P}_-)\ \dd\sigma_R\ +\ 
             (1-{\rm P}_+)(1-{\rm P}_-)\ \dd\sigma_L\right],
\label{sigmapol}
\eeq
where $\sigma_{R/L}$ corresponds to the cross section for electrons and
positrons with equal right/left-handed helicity. Chirality conservation
suppresses opposite helicities. Table~\ref{tab10} summarizes some extreme
cases. If both beams are fully polarized, ${\rm P}_+={\rm P}_-=\pm 1$, the 
ratio $r$ is the same as for (${\rm P}_\pm=0$, ${\rm P}_\mp=\pm 1$), 
respectively, but the cross sections are twice as much (\ref{sigmapol}),
which results in a higher statistical significance of the CP signal. From 
comparison of Table~\ref{tab9} (right columns) with Table~\ref{tab10}
is clear that left-handed polarized beams enhance the sensitivity to 
CP-violating effects.

\begin{table}[ht]
\begin{center}
\begin{tabular}{|c|r|r|r|}
\hline
\multicolumn{4}{|c|}{${\rm P}_\pm=0$, ${\rm P}_\mp=-1$; 
                $\sigma_{t\bar t}=0.707$ pb} \\
\hline
${\cal O}$ & $b-b$ & $\ell-b$ & $\ell-\ell$ \\
\hline \hline
$\hat{A}_1$   &
\ \ 0.359       &
$-0.298$       &
$-0.667$        \\
$\hat{A}_2$       &
0.397       &
$-0.254$       &
$-0.952$        \\
$\hat{T}_{33}$&
0.065       &
$-0.214$       &
0.383        \\
$\hat{Q}_{33}$&
0.708       &
$-0.488$       &
$-1.848$        \\
\hline
\end{tabular}
\begin{tabular}{|c|r|r|r|}
\hline
\multicolumn{4}{|c|}{${\rm P}_\pm=0$, ${\rm P}_\mp=+1$; 
$\sigma_{t\bar t}=0.355$ pb} \\
\hline
${\cal O}$ & $b-b$ & $\ell-b$ & $\ell-\ell$ \\
\hline \hline
$\hat{A}_1$   &
$0.016$       &
$-0.009$       &
$-0.053$        \\
$\hat{A}_2$       &
0.134       &
$-0.099$       &
$-0.468$        \\
$\hat{T}_{33}$&
$-0.003$       &
$0.006$       &
$-0.020$        \\
$\hat{Q}_{33}$&
0.210       &
$-0.145$       &
$-0.555$        \\
\hline
\end{tabular}
\end{center}
\caption{\sf The same as in Table~\ref{tab9} assuming the complete one-loop 
cross section for $e^+e^-\to t\bar t$ and longitudinal polarizations for
one of the $e^\pm$ beams, ${\rm P}_\pm$. \label{tab10}}
\end{table}

\bigskip

Concerning the experimental reach of this analysis,
the statistical significance of the signal of CP violation is given
by $N_{SD}=|r|\sqrt{N}$ with $N=\epsilon{\cal L}\sigma_{t\bar t}\mbox{BR}
(t\to a)\mbox{BR}(\bar t\to\bar c)$ where $\epsilon$ is the detection
efficiency and ${\cal L}$ the integrated luminosity of the collider.
The branching ratios of the $t$ decays are BR $\simeq 1$ for the $b$ channel
and BR $\simeq 0.22$ for the leptonic channels ($\ell=e,\mu$).
At $\sqrt{s}=500$ GeV the total cross section for $t$-pair production is
$\sigma_{t\bar t}\simeq 0.5$ pb. Assuming a LC integrated luminosity ${\cal L}
\simeq 500$ fb$^{-1}$ and a perfect detection efficiency, one
gets $\sqrt{N}\simeq 500, 235, 110$ for the channels $b-b$, $\ell-b$,
$\ell-\ell$, respectively. With these statistics, values of $|r|\sim 
2\times 10^{-3}, 4\times 10^{-3}, 9\times 10^{-3}$ for the channels above
would be necessary to achieve a 1 s.d. effect. Such values can be hardly reached 
in the context of the MSSM for the given luminosity, even for polarized beams, 
as Tables~\ref{tab9} and \ref{tab10} show.

\section{Conclusions}

In summary, the top-quark is an excellent laboratory to test the MSSM. If SUSY
is realized in nature around the electroweak scale, the top-quark phenomenology
differs significantly from the SM one. The various top-quark observables can be well investigated at TESLA.
The production cross-section receives radiative corrections that can be used to
test the MSSM, or to derive bounds on the masses of the various
sparticles. Also, in the MSSM there exist new sources of CP-violation effects,
giving rise to specific CP-violating asymmetries in top-pair production, which,
however, might be too small for being observable with the currently expected
luminosity.  The top-quark decay modes can differ significantly from the SM
ones; of special importance is the $t\to H^+b$ decay mode. Even in the case that
the new particles of the MSSM are heavier than the top-quark, their virtual
quantum effects influence the top-quark standard decay mode. Moreover, the LC
presents a great sensitivity to the possible rare decays. In the MSSM some of
these decays are strongly enhanced with respect to the SM expectations, and
could be detected.

\section*{Acknowledgments}
This work has been partially supported by the Deutsche Forschungsgemeinschaft,
by CICYT under projects AEN99-0766 and AEN96-1672 and by Junta de 
Andaluc{\'\i}a under project FQM-101.

\end{document}